\documentclass[12pt,a4paper]{article} 

\usepackage{makecell}
\usepackage{multirow}
\usepackage{booktabs}
\usepackage{hhline}
\usepackage{colortbl}
\usepackage{color}
\usepackage{rotating}
\usepackage{tabularray}
\usepackage[export]{adjustbox}
\usepackage{graphicx}
\usepackage{caption} 
\usepackage{amssymb} 
\usepackage{amsmath}
\usepackage{setspace}

\usepackage[
  colorlinks=true,
  linkcolor=blue,
  citecolor=blue,
  urlcolor=blue
]{hyperref}
\usepackage{cleveref}
\usepackage{authblk}

\usepackage{epsfig}
\usepackage{graphics}
\usepackage{subfigure}
\usepackage{a4wide}
\usepackage{amsmath}
\usepackage{diagbox}
\usepackage[normalem]{ulem}

\graphicspath{{Fig/}}

\title{\Large{\textbf{ Hamiltonian approach to isospinning ${\mathbb C}P^2$ solitons}}}

\author{Sergei Antsipovich\footnote{E-mail: srh.antsp@gmail.com}}

\affil{Department of Theoretical Physics and Astrophysics,\\
Belarusian State University, Minsk 220004, Belarus}

\providecommand{\keywords}[1]
{
  \medskip
  \textbf{{Keywords:}} #1
}

\begin{document}

\maketitle

\abstract{\noindent
Isorotating ${\mathbb C}P^2$ Q-solitons in 2+1 dimensions were studied. Hamiltonian formalism as a more physically meaningful yet fairly demanding approach was adopted during the investigation, which helped to exclude unobservable parameters such as angular frequencies and Lagrangian. This approach also highlighted the non-topological nature of the stabilization mechanism and revealed a number of similarities between well-known $U(1)$ Q-balls and ${\mathbb C}P^2$ isospinning solitons, thus rendering the latter a suitable extension of the former for the case of higher Lagrangian symmetry group and paving the way for further ${\mathbb C}P^N$ generalizations. Due to the peculiarities of the model, numerical optimisation algorithms were chosen to obtain the solutions.

}


\newpage

\section{Introduction \label{sec:introduction}}

Nowadays Q-solitons are still ones of interest \cite{MareikeThesis, Cantara, AANS, Shnir, Zhou}. The core principle is the phenomenon of isorotation in the target space, which helps to stabilize the solutions in the absence of terms of higher nonlinearity. So far, the most well-known investigated cases are one-parametric U(1) and O(3) models \cite{Ward, MareikeThesis}. Studies on multi-parametric Q-solitons, in turn, are usually represented by models with a number of complex scalar fields coupled in the Lagrangian \cite{FORGACS, Loginov:2019rwz}, yet some special cases of genuine non-abelian Q-balls have been studied as well \cite{AXENIDES1998190, SAFIAN1988498}. 
\par
The ${\mathbb C}P^N$ models, however, appear to allow one to extrapolate the principles of U(1) Q-balls on Lagrangians with higher unitary Lie groups symmetries.
\par
Present paper may be regarded as a supplementary extension of a recent major study dedicated to ${\mathbb C}P^2$ solitons \cite{AANS}. An alternative approach to numerical investigation made it possible to cover a case with $N=1$ topological charge in more detail. The method used in the aforementioned article is based on Newton-Raphson algorithm and was applied to a second-order Euler-Lagrange ODE system. It was shown that in case of unit topological charge profile functions may admit a manual redefinition of boundary conditions for the numerical method to converge. Such a redefinition is safe due to a certain degree of freedom existing when defining the vacuum in zero (see Subsec.~\ref{subsubsec:analiticity}).
\par
One of present paper's novelties is a different numerical method (Subsec.~\ref{subsec:methods}) based on energy minimization \cite{Hale}. Along with that, two different potentials were analysed and their solutions compared. The solitons obtained are found to be stable both with respect to Derrick's scaling argument and towards the decaying into free particles, see Sec.~\ref{sec:existence_and_stability}.

\section{Model \label{sec:model}}

Let us consider a Lagrangian density of ${\mathbb C}P^2$ NL-$\sigma$ model \cite{AASS, AAGNS} in a flat 2+1 space-time with a metric signature $\left ( 1,-1,-1 \right )$. In this model the Dirichlet term alone possesses several symmetries, to be specific global SU(3) and local abelian U(1) \cite{Zakrzewski}. However, to obtain stable Q-solitons, one should introduce an additional symmetry breaking potential term $\mathcal{U}$, which will stabilize the solution. Altogether, one gets
\begin{equation}\label{LagrangianN}
\mathcal{L}=\frac{1}{4}\textup{Tr}\left ( \partial_{\mu } \mathcal{\mathfrak{n}} \: \partial^{\mu } \mathcal{\mathfrak{n}}\right)-\mathcal{U},
\end{equation}

where $\mathfrak{n}=n^{a}\lambda _{a}$ ($a=1,2,...,8$). Here, the real scalars $n_{a}$ represent components of so-called ``colour field" \cite{Kondo}, \cite{AASS} with a basis composed of Gell-Mann matrices $\lambda_{a}$.
Since a sigma-model is considered, the field is subject to the following constraints \cite{Kondo2}, \cite{AAGNS}: 
% $\textup{Tr}\! \left ( \mathfrak{n}\mathfrak{n} \right )=\frac{8}{3} $ and $ \left\{ \mathfrak{n},\mathfrak{n} \right\}=\frac{4}{3}\mathfrak{n}+\frac{16}{9}\textbf{1}_{3}$.
\begin{equation*}
    \textup{Tr}\! \left ( \mathfrak{n}\mathfrak{n} \right )=\frac{8}{3} \;\text{ and } \;n^{a}=\frac{3}{2}d_{abc}n^{b}n^{c}, 
\end{equation*}
where completely symmetric coefficients $d_{abc}=\frac{1}{4}\textup{Tr}\left ( \lambda_{a} ,\left\{ \lambda_{b} ,\lambda_{c} \right\} \right )$.

\par
Yet, there is another formalism available \cite{DinZakrzewski1, DinZakrzewski2}, which utilises homogeneous coordinates on the ${\mathbb C}P^2$ per se, namely $ Z\in \left\{z\in  \mathbb{C}^{\, 3}|\: z^{\dagger } z=1\right\}$. Colour field components, in turn, are parameterized as follows 
\begin{equation}\label{n_def}
    n^{a}=Z^{\dagger }\lambda _{a}Z,
\end{equation}
so that Lagrangian density casts into 

\begin{equation}\label{LagrangianZ1}
    \mathcal{L}=2 \left( D_{\mu } Z\right)^{\dagger }D^{\mu } Z - \mathcal{U}.
\end{equation}

Here the covariant derivatives $D_{\mu}=\partial _{\mu}-\textit{i}A_{\mu}$ are of composite gauge type \cite{Bitar}, with the gauge field $A_{\mu}=-\textit{i}Z^{\dagger }\partial _{\mu}Z$.

\par

The corresponding energy-momentum tensor density can be expressed as follows:

\begin{equation}\label{EMT}
     T_{\; \; \nu}^{\mu}=2 \left( \left ( D^{\mu }Z \right )^{\dagger }D_{\nu }Z+\left ( D_{\nu }Z \right )^{\dagger }D^{\mu }Z \right ) -\delta_{\; \; \nu}^{\mu} \mathcal{L}
\end{equation}

where $\mathcal{L}$ is the Lagrangian density \eqref{LagrangianZ1} and $\delta^{\mu}_{\nu}$ is Kronecker's delta.

\par

To construct Q-lumps one must first derive Noether charges of the model. Using homogeneous coordinate formalism \eqref{LagrangianZ1} and formula for current of complex vector field one may obtain conserving currents' densities

\begin{equation}\label{current}
j_{\mu }^{\alpha}=\textit{i}\left ( \frac{\partial\mathcal{L} }{\partial \left ( \partial_{\mu } Z \right )}\Lambda_{\alpha} Z-Z^{\dagger }\Lambda_{\alpha} \frac{\partial\mathcal{L} }{\partial \left ( \partial_{\mu } Z^{\dagger } \right )} \right ),
\end{equation}

where $\Lambda_{\alpha}$ is a generator of the corresponding symmetry of the Lagrangian. 
\par
Speaking of the symmetries, it may be shown that considering potentials of the form $\mathcal{U}=\mathcal{U}\left(n^{3},n^{8}\right)$, the global symmetry of the overall Lagrangian is broken down to the maximal torus or Cartan subgroup of $SU\!\left(3\right) \textup{ mod } U(1)$. Indeed, global transformations $Z\mapsto gZ$ with $g\in \left\{ h\in U\!\left ( 3 \right )\,|\,\left [ h,\lambda_{8} \right ]=0\right\}$ leave $\left ( n^{3},n^{8} \right )$ intact, which is clear from \eqref{n_def}. As for the specific values of the generators, in this paper the following set will be used:
\begin{equation}\label{tau_generators}
    \tau_{1}=\textup{diag}\left ( 0,1,0 \right)\;\text{ and  }\; \tau_{2}=\textup{diag}\left ( 0,0,1 \right).  
\end{equation}
The model also possesses a topological charge invariant under the \textit{local} $U(1)$ transformations, which is given by the formula:
\begin{equation}\label{topChargeIntegral}
Q_{\textup{top}}=-\frac{\textit{i}}{2\pi }\int \textup{d}^{2}x\: \varepsilon_{jk}(D_{j}Z)^{\dagger }D_{k}Z.
\end{equation}

\section{Ansatz \label{sec:ansatz}}

Speaking of Q-solitons as of the lowest-energy field configurations for fixed charges, it is necessary to choose an appropriate form of the solution to seek for. \par
Without loss of generality one may parametrize a unimodular complex vector as $Z=e^{\textit{i}\,\Theta }X$, with $\Theta (x)$ being a $3\!\times\!3$ real diagonal matrix and $X(x)$ as a real vector on unit $S^{2}$, i.e.
\begin{equation}\label{Xvec}
       X\in \left\{\chi\in  \mathbb{R}^{ 3}\mid\chi^{T} \chi=1\right\}.
\end{equation}
Then, for each of the generators $\tau_{\alpha}$ we have conserving Noether charges \eqref{current}, densities of which $\rho_{\alpha}\equiv j_{0}^{\alpha}$ can be explicitly written in an above chosen ansatz: 
\begin{equation}
 \frac{1}{4} \rho_{\alpha}= X^{T}\left ( \partial _{t}\Theta  \right )\tau_{\alpha } X
   -\left ( X^{T} \tau_{\alpha } X \right )\left ( X^{T}\partial _{t}\Theta X \right)
\end{equation}

The minimal energy solution with fixed charges $Q_{1}$ and $Q_{2}$ can be sought by introducing Lagrange multipliers $\omega_{1}$ and $\omega_{2}$, respectively. Thus, constrained energy functional takes form
\begin{equation}
    E_{Q}=E-\omega _{1}\left ( Q_{1}-\int_{V_{2}}\! \rho_{1}  \right )-\omega _{2}\left ( Q_{2}-\int_{V_{2}}\! \rho_{2}  \right ),
\end{equation}
where, according to \eqref{EMT}, $E=\int_{V_{2}} \! T_{00}$ and $\int_{V_{2}}$ is an integral by 2-dimensional volume.
In a more explicit form the above expression can be written as
\begin{equation}
\begin{aligned}\label{E_constr}
  E_{Q}&=\omega_{1} Q_{1}+ \omega_{2} Q_{2} +2\!\int _{V_{2}} \left ( D_{j}Z \right )^{\dagger }D_{j}Z+\int _{V_{2}}\mathcal{U}\left(X\right) + 2\!\int _{V_{2}}\partial_{t}X^{T}\partial_{t}X \\
& + 2\!\int _{V_{2}}\uline{X^{T}\left (  \omega_{1}\tau_{1}+\omega_{2}\tau_{2} - \partial _{t}\Theta \right )^2 X}-2\!\int _{V_{2}}\uline{ \bigl( X^{T}\left ( \omega_{1}\tau_{1}+\omega_{2}\tau_{2} -\partial _{t}\Theta \right ) X  \bigr)^{2}}\\
&-2\!\int _{V_{2}}X^{T}\left (  \omega_{1}\tau_{1}+\omega_{2}\tau_{2} \right )^2 X+2\!\int _{V_{2}}\bigl( X^{T}\left ( \omega_{1}\tau_{1}+\omega_{2}\tau_{2} \right ) X \bigr)^{2},
\end{aligned}
\end{equation}

where $j$ index iterates through spatial coordinates. One more thing that might be of help is spherical rearrangement theorem \cite{COLEMAN1985263}. It implies that minimal-energy solutions should be rotationally invariant in order to minimize gradient energy. Thus, it is relevant to use polar coordinates when considering the gradient term:

\begin{equation}
\begin{aligned}\label{grad_term}
    \left ( D_{j}Z \right )^{\dagger }D_{j}Z&=\partial_{r}X^{T}\partial_{r}X+\dashuline{X^{T }\left ( \partial _{r}\Theta  \right ) ^{2}X}-\dashuline{\left( X^{T }\partial _{r}\Theta   X  \right)^{2}}\\
    &+\frac{1}{r^{2}}\left (\partial_{\theta}X^{T}\partial_{\theta}X+\dotuline{X^{T }\left ( \partial _{\theta}\Theta  \right ) ^{2}X}-\dotuline{\left ( X^{T }\partial _{\theta}\Theta   X \right )^{2}}\right ).
\end{aligned}
\end{equation}
Similarly to \cite{COPPERSMITH1997201}, it can be shown that with $A^{\dagger}=A$ the inequality $Z^{\dagger}A^2Z\geqslant \left(Z^{\dagger}AZ\right)^{2}$ holds, thus underlined combinations above are always non-negative.
A number of positive definite terms in \eqref{E_constr} and \eqref{grad_term} may be cancelled by setting  $\partial _{t}X=\partial _{\theta}X=0$ and $ \partial_{r}\Theta =0$. It also becomes clear, that $\Theta =\left ( \omega_{1}\tau_{1} +\omega_{2}\tau_{2} \right )t+\Phi(\theta )$, where remaining winding function $\Phi(\theta+2\pi)=\Phi(\theta)$ is periodical. Since the objective of this work also includes studying angular momentum, it is necessary to keep $ \partial_{\theta}Z\not\equiv  0$. The minimal energy satisfying such condition can be achieved by uniform winding function, which sets $\Phi\!\left ( \theta  \right )=\left ( k_{1}\tau_{1}+k_{2}\tau_{2} \right )\theta $, with winding numbers $k_{1,2}\in \mathbb{Z}$.
\par
As it is now only spherically-symmetric Q-ball solutions are considered, one may factorize the field acknowledging its temporal and angular periodicity. This can be performed the following way:
\begin{equation}\label{factor}
     Z\!\left ( x \right )=e^{\textit{i}\,\Omega t}e^{\textit{i} K\theta }X(r).
\end{equation}

Here the temporal and spatial rotation generators are hermitian, due to being a linear combination of Cartan subalgebra elements \eqref{tau_generators}, namely

\begin{equation}\label{K_Omega}
    K =k_{1} \tau_{1}+k_{2} \tau_{2}\; \text{ and }\; \Omega  =\omega _{1} \tau_{1}+\omega _{2} \tau_{2}.
\end{equation}

With a given ansatz \eqref{factor} and constraint \eqref{Xvec}, the correct asymptotic behaviour and topology of solutions can be provided via following parametrization:
\begin{equation}\label{X_profile_funs}
X=\left ( \textup{cos}\! \left ( F \right ),\,
\textup{sin} \! \left ( F \right ) \textup{cos} \! \left ( G \right ),\,
\textup{sin} \! \left ( F \right ) \textup{sin} \! \left ( G \right ) \right )^{\textup{T}},
\end{equation}

where $F(r)$ and $G(r)$ are radial profile functions taking values on $[0, \frac{\pi}{2}]$.

\section{Lagrangian formulation \label{sec:lagrangian_form}}

Total Lagrangian of density \eqref{LagrangianZ1} may be regarded as being composed of two distinct terms of different nature, namely  
\begin{equation}\label{two_dist_terms}
    \mathcal{L}_{2}=2\left ( D_{j}Z \right )^{\dagger }D_{j}Z\; \text{ and } \;\mathcal{U}_{\textup{eff}}=\mathcal{U}-\mathcal{L}_{0},
\end{equation}

where $\mathcal{U}_{\textup{eff}}$ is an effective potential consisting of proper potential $\mathcal{U}$ and $\mathcal{L}_{0}$, which is a kinematic part of Lagrangian density \eqref{LagrangianZ1}. It may be rewritten in terms of the ansatz \eqref{factor} as
\begin{equation}\label{L0density}
    \mathcal{L}_{0}=2 \left ( D_{t}Z \right )^{\dagger } D_{t}Z=2\Bigl( X ^{T}\Omega ^{2} X- \bigl( X^{T}\Omega  X\bigr)^{2}\Bigr).
\end{equation}

In the same manner, the Noether charges' densities \eqref{current} $j^{\alpha}_{0}\equiv \mathcal{Q}_{\alpha}$ are expressed as follows:
\begin{subequations}\label{charge_both_density}
\begin{align}
\begin{split}\label{charge1_density}
\mathcal{Q}_{1}=4\Bigl(X^{T}\Omega \tau_{1} X-\bigl(  X^{T}\tau_{1}  X\bigr) \bigl(X^{T}\Omega  X\bigl)\Bigr),
\end{split}\\
\begin{split}\label{charge2_density}
\mathcal{Q}_{2}=4\Bigl(X^{T}\Omega \tau_{2} X-\bigl(X^{T}\tau_{2} X\bigr) \bigl( X^{T}\Omega  X\bigl)\Bigr).
\end{split}
\end{align}
\end{subequations}

It is not hard to see that now using \eqref{K_Omega} both the Lagrangian and Noether charges may be formulated in terms of "inertia tensor" density $\mathcal{I}$, the components of which are given below 
\begin{subequations}\label{I_ij}
\begin{align}
\begin{split}    
\label{I_ij_begin}
 \mathcal{I}_{11}=4\Bigl( X ^{T} \tau_{1}^{2} X-\bigl( X^{T}\tau_{1}X\bigr)^{2} \Bigr)
 \end{split}\\
 \begin{split}
 \mathcal{I}_{12}=4\Bigl( X^{T} \tau_{1} \tau_{2} X-\bigl( X^{T}\tau_{1} X\bigr)\bigl(  X^{T}\tau_{2} X\bigr) \Bigr)
 \end{split}\\
 \begin{split}
\mathcal{I}_{22}=4\Bigl(X^{T} \tau_{2}^{2} X-\bigl( X^{T}\tau_{2}  X\bigr)^{2} \Bigr).\label{I_ij_end}
\end{split}
\end{align}
\end{subequations}

By introducing new notations for $\vec{\omega }=\left ( \omega _{1}, \omega _{2} \right )^{T}$ and for $\vec{Q }=\left ( Q_{1}, Q_{2} \right )^{T}$, along with the integrals of \eqref{I_ij} components $I_{\alpha\beta}=\int\!d^{2}x\,  \mathcal{I}_{\alpha\beta}$ ($\alpha,\beta =1,2$), the total kinematic part of the Lagrangian $L_{0}=\int d^{2}x\mathcal{L}_{0}$ and total Noether charges $Q_{\alpha}=\int\!d^{2}x\,  \mathcal{Q}_{\alpha}$ are now casted into a rather concise form:

\begin{equation}\label{L0Q}
 L_{0}=\frac{1}{2} \vec{\omega}^{\: T}\! I\vec{\omega} \;\;\text{ and }\;\;   \vec{Q}=I\vec{\omega }.  
\end{equation}
Such a form resembles classical dynamics of rigid body rotational motion \cite{Shnir, MareikeThesis} and will be used further.

Now, having ascertained \eqref{L0density} to be a stationary function, let's consider its vacua.
Pretty much as the potential itself, the effective potential must tend to zero as the field approaches vacuum state in the infinity. Retaining $\omega_{1}, \omega_{2}$ arbitrary, this requirement may be achieved by assuming $\underset{Z\to Z_{\infty}}{\textup{lim}}\mathcal{I}_{\alpha\beta}=0$.
\par
As $\tau_{\alpha}$ have been shown to be diagonal (see Sec.~\ref{sec:ansatz}), it is clear that such a condition is satisfied when 
\begin{equation}\label{basis_cond}
  Z_{\infty}\;\textup{mod}\; U(1)\in \left\{ \boldsymbol{e}_{1},\boldsymbol{e}_{2},\boldsymbol{e}_{3}\right\},  
\end{equation}
i.e. when the vacuum is a vector of $\mathbb{R}^3$ standard basis.
\par
Summing up, the solutions themselves may be found in a form of profile functions \eqref{X_profile_funs} as solutions of a nonlinear second-order Euler-Lagrange ODEs system  
\begin{equation}\label{full_EL}
    \left\{\begin{matrix}
\frac{1}{r}\partial_{r}\!\left ( r\left ( \frac{\partial \mathcal{L}}{\partial \left ( \partial_{r} F \right )} \right ) \right )-\partial_{F}\mathcal{L}=0 \\
\frac{1}{r}\partial_{r}\!\left ( r\left ( \frac{\partial \mathcal{L}}{\partial \left ( \partial_{r} G \right )} \right ) \right )-\partial_{G}\mathcal{L}=0,
\end{matrix}\right.
\end{equation}
with appropriately chosen boundary conditions $F(0), G(0)$ and $F(\infty ), G(\infty )$.
Such an approach was utilized in \cite{AANS}, but is not the subject of the present article.
\par
To conclude, there is another integral of motion which can be expressed keeping with the same approach as in \eqref{L0Q}. This quantity is the angular momentum, density of which can be obtained from the EMT \eqref{EMT} using Eqs.~\eqref{factor}. An intermediate form is expressed as
\begin{equation*}
    T_{\mathit{0} \theta }=4\Bigl( X^{T}\Omega K X-\bigl( X^{T} K X \bigr)\bigl(X^{T} \Omega X\bigr) \Bigr),
\end{equation*}
thus a clear similarity reveals that it as well can be written in terms of Noether charges by simplifying $K$ generator, see~\eqref{K_Omega}. Indeed, the previous relation decomposes into two terms, proportional to densities of both the charges \eqref{charge1_density}, \eqref{charge2_density}: 
\begin{equation}\label{angular_m}
    J=\int T_{\mathit{0} \theta }\: d^{2}x=k_{1}Q_{1}+ k_{2}Q_{2}=\vec{k}\cdot\vec{Q}.
\end{equation}
Hence, angular momentum appears to be a linear combination of two Noether charges with quantum numbers $k_{1,2}$ taken as weights.

\section{Legendre transform. Hamiltonian formulation \label{sec:legendre}} 

The model also possesses a more appeling from the physical point of view hamiltonian formulation. In order to investigate it one needs to decompose the Lagrangian into two distinct terms, namely $L_{0}$, corresponding to the stationary kinematic part and $M_{N}$ being a static ${\mathbb C}P^2$ energy \cite{AASS}, more specifically $M_{N}=\int d^{2}x \left ( 2 \left ( D_{j}Z \right )^{\dagger } D_{j }Z-\mathcal{U} \right ) $. Thus, one gets
\begin{equation}
    L=L_{0}-M_{N}
\end{equation}
Since \eqref{L0density} is strictly nonnegative, it can be concluded that inertia tensor \eqref{I_ij_begin}-\eqref{I_ij_end} is a nonnegative-semidefined matrix, same can be said about its inverse. Assuming the invertibility of $I$, from \eqref{L0Q} it is also clear that $L_{0}=\frac{1}{2}\vec{Q}^{T}I^{-1}\vec{Q}$. 
\par
This result can also be obtained from the corresponding Legendre transform. Indeed, definition of a new canonical variable,
\begin{equation}\label{canonvar}
   \frac{\partial L}{\partial \vec{\omega}}= \vec{Q},
\end{equation}
leads to a standard expression of the energy functional 
\begin{equation}\label{hamiltonian}
    H=\vec{\omega}\cdot\frac{\partial L}{\partial \vec{\omega}}-L=\vec{\omega}\cdot\vec{Q}-L=L_{0}+M_{N},
\end{equation}
what, in turn, is in agreement with the value of $T_{00}$ component of EMT, see Eq.~\eqref{EMT}.
\par
Beyond all the known implications of the Legendre transform, the one is worth looking at in more detail here. Namely, the solutions can be obtained either from pseudoenergy $F=-L$ extremization under fixed frequencies, or, equivalently, from Hamiltonian minimization under fixed charges. Indeed, using \eqref{L0Q} one may get
\begin{equation*}
    \begin{aligned}
        \delta H|_{\vec{Q}}&=\delta L_{0}|_{\vec{Q}}+\delta M_{N}=\frac{1}{2}\vec{Q}^{T}\delta \!\left ( I^{-1} \right )\vec{Q} +\delta M_{N}=\\
&-\frac{1}{2}( \vec{Q}^{T}I^{-1} )\delta \! I( I^{-1}\vec{Q} )+\delta M_{N}=-\frac{1}{2}\vec{\omega }^{T}\delta \!I\vec{\omega }+\delta M_{N}=-\delta L_{0}|_{\vec{\omega}}+\delta M_{N}=-\delta L |_{\vec{\omega}}.
    \end{aligned}
\end{equation*}
Such an approach enables us to switch to a more physical description of the system, sticking to observable parameters like energy and charges and distracting from unobservable ones -- internal frequencies and Lagrangian.

\section{Existence and stability \label{sec:existence_and_stability}}
\subsection{Choice of potential}
\begin{figure}
	\centering
	\adjincludegraphics[trim={{.0\width} {.0\width} {.0\width} {.0\width}},clip, width=0.34\textwidth]{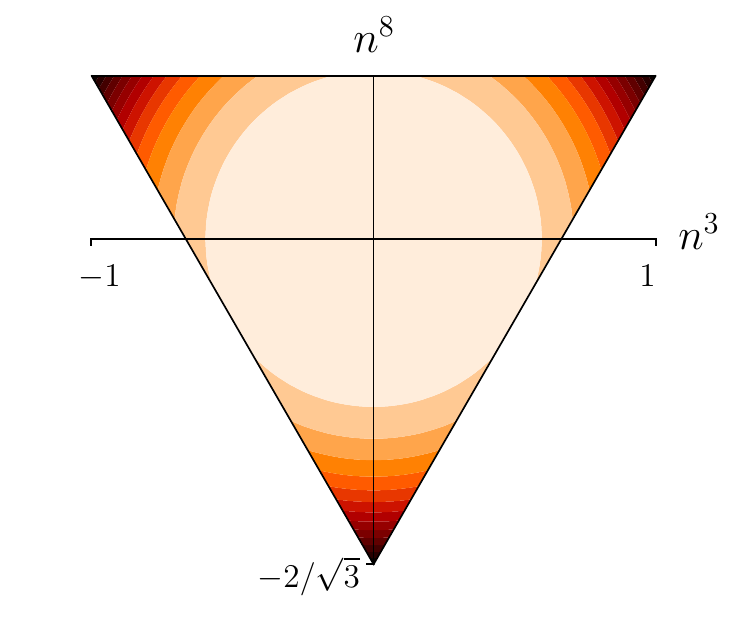}%
	\adjincludegraphics[trim={{.0\width} {.0\width} {.4\width} {.0\width}},clip, width=0.137\textwidth]{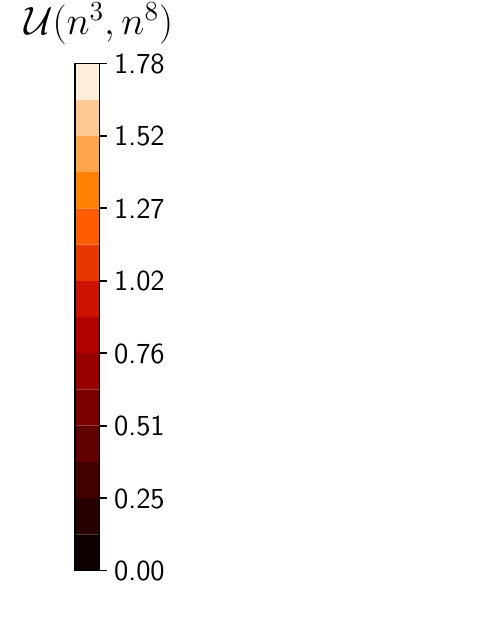}%
    \adjincludegraphics[trim={{.0\width} {.0\width} {.0\width} {.0\width}},clip, width=0.34\textwidth]{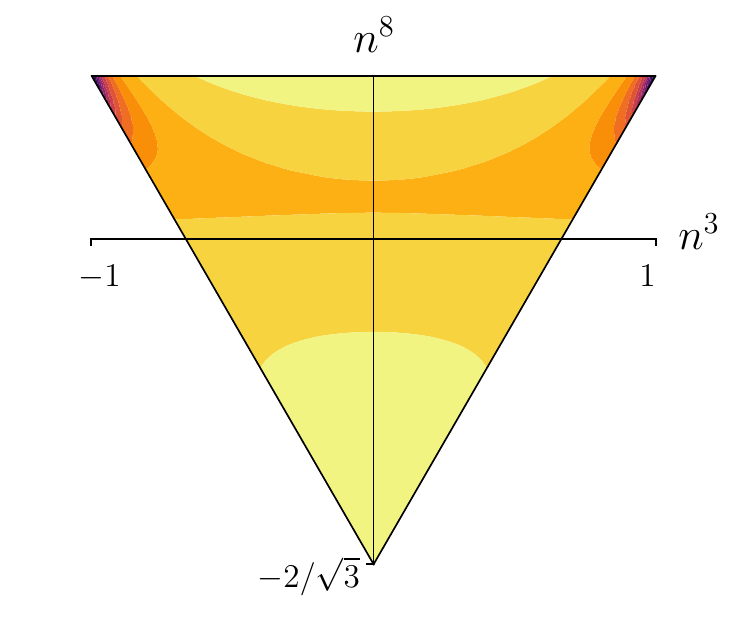}%
	\adjincludegraphics[trim={{.0\width} {.0\width} {.4\width} {.0\width}},clip, width=0.137\textwidth]{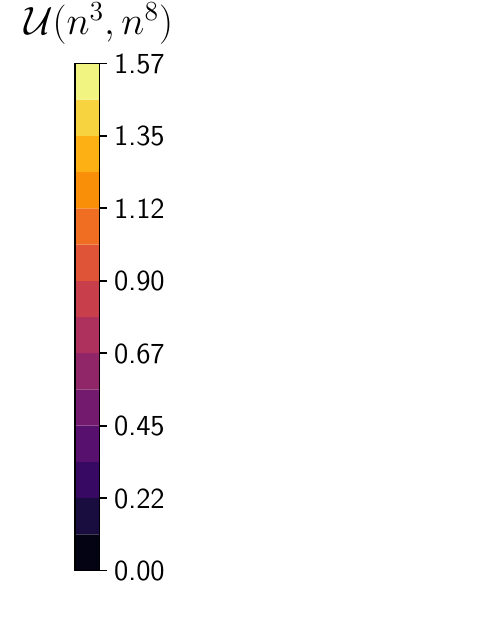}%
	\caption{Three-vacuum potential (left) and two-vacuum potential (right).}
	\label{both_potentials}
\end{figure}

The possibility of existence of stable solitonic solutions is heavily dependent on the choice of the potential. In the given paper two types of $U(1)\!\times\! U(1)$-symmetrical functions were considered:
\begin{itemize}
    \item Two-vacua potential (see Fig.~\ref{both_potentials}, right), which had already been partly analysed in the previous work \cite{AANS}.
\begin{equation}\label{U_2_vac}
    \mathcal{U}=\mu ^{2}\,  \textup{arctan}\! \left (5\; \mathcal{W}\!\left ( n^{3}, n^{8} \right ) \right ),
\end{equation}
where explicit form of $\mathcal{W}$ is
\begin{equation*}\label{potW}
\mathcal{W}(n^3, n^8)=(2+ 3 n^3 + \sqrt 3 n^8)(2- 3 n^3 + \sqrt 3 n^8)(n^8)^2 + (1-\sqrt 3 n^8).
\end{equation*}
It may be seen to be symmetric under the transformation $\left(n^{3},n^{8}\right)\to\left(-n^{3},n^{8}\right)$. The vacua of this potential are $\bigl( 1,\frac{1}{\sqrt{3}} \bigr)$ and $\bigl(- 1,\frac{1}{\sqrt{3}} \bigr)$.
    \item Three-vacua potential (see Fig.~\ref{both_potentials}, left)
\begin{equation}\label{U_3_vac}  
 \mathcal{U}=\mu^{2}\left ( \frac{16}{9}-\left ( (n^{3})^2+(n^{8})^2 \right )^{2}\right ).
\end{equation}
It possesses a higher symmetry of $D_{3}$, which is the dihedral group of the equilateral triangle.
\par
It can be indeed seen that \eqref{LagrangianN} with \eqref{U_3_vac}
is invariant under the action of following generating elements 
\begin{subequations}
\begin{align}
\begin{split}\label{n_reflection}
 \textstyle\left ( n^{3},n^{8} \right )\xrightarrow{S}\left (- n^{3},n^{8} \right )
 \end{split}\\
 \begin{split}\label{n_rotation}
\textstyle\left ( n^{3},n^{8} \right )\xrightarrow{R}\left (-\frac{1}{2}n^{3}+\frac{\sqrt{3}}{2}n^{8},-\frac{\sqrt{3}}{2}n^{3}-\frac{1}{2}n^{8}\right ),
 \end{split}
\end{align}
\end{subequations}
being a reflection over $n_{3}$-axis and clockwise $2\pi/3$ rotation around $(0,0)$, respectively.
The vacua of this potential are $\bigl( 1,\frac{1}{\sqrt{3}} \bigr),\bigl(- 1,\frac{1}{\sqrt{3}} \bigr)\text{ and }\bigl( 0,-\frac{2}{\sqrt{3}} \bigr)$.
\end{itemize}
For both potentials $\mu$ is a bare mass parameter. It is also worth noting that all the vacua of the above potentials are in coherence with the condition imposed in \eqref{basis_cond}. 

\subsection{Stability
\label{subsec:stability}}

There are several main stability indicators that are used to study Q-solitons. Among them, Derrick's theorem \cite{Ward}, quantum stability condition \cite{LeePang}, classical stability criterion, and stability towards fission \cite{Hamada}. The former two are utilized in the present article.

\subsubsection{Derrick's theorem
\label{subsubsec:derricks_theorem}}

Let us consider an integral of Lagrangian density \eqref{LagrangianZ1}. It may be shown that transformation of spatial coordinates $\boldsymbol{x}\mapsto \lambda \boldsymbol{x}$ scales total values of \eqref{two_dist_terms} as $L_{2}\mapsto L_{2}$ and $U_{\textup{eff}}\mapsto\lambda^{2}U_{\textup{eff}}$ respectively.
\par
Since solitons are stationary points of the Lagrangian, the solutions must satisfy $\partial _{\lambda}L\left ( \lambda  \right )\big|_{\lambda =1} =0$, or, explicitly,

\begin{equation}\label{virial}  
0=U_{\textup{eff}}=\int_{V_{2}}\left(\mathcal{U}-\mathcal{L}_{0}\right)\Rightarrow U=L_{0}.
\end{equation}

Hereinafter, this condition will be referred to as virial relation.
It also defines a ``lower bound" on $\vec{\omega}$, or a subset to which $(\omega_{1},\omega_{2})$ must belong in order for the soliton to be stable. This subset is

\begin{equation}\label{lower_bound}
\begin{aligned}
    \mho_{-}\!\left( \mu \right )
     &= \bigl\{\left ( \omega_{1},\omega_{1} \right )\mid\exists \left ( F_{1},G_{1} \right ):\mathcal{L}_{0}\left (F_{1},G_{1},\omega_{1},\omega_{2} \right )>\mathcal{U}\left ( F_{1},G_{1},\mu \right )\\ 
     & \;\text{ and }\;\exists \left ( F_{2},G_{2} \right ):\mathcal{L}_{0}\left (F_{2},G_{2},\omega_{1},\omega_{2} \right )<\mathcal{U}\left ( F_{2},G_{2},\mu \right ) \bigr\},
\end{aligned}
\end{equation}

where $\mathcal{L}_{0}$ and $\mathcal{U}$ are formulated as functions of $F,G$ \eqref{X_profile_funs}. Resolving this domain, in general, is possible only numerically, yet does not involve searching for the solutions per se \cite{AANS}.

\subsubsection{Quantum stability
\label{subsubsec:quantum_stability}}

Also known as absolute stability or stability towards fission into free quantum particles \cite{Tsumagari2009ThePO, Gulamov:2013cra}, it is derived by considering small perturbations of the field near infinity's vacuum. In terms of profile functions \eqref{X_profile_funs} it can be given as $\underset{r\rightarrow \infty }{\textup{lim} }F\!\left ( r \right )= F\!\left ( \infty  \right )+f\!\left ( r \right )$
 and 
$\underset{r\rightarrow \infty }{\textup{lim} }G\!\left ( r \right )= G\!\left ( \infty  \right )+g\!\left ( r \right )$, where $\left | f\right |,\left | g\right |\ll 1$. 
Now, considering the first order of fluctuations' smallness, the Eqs.~\eqref{full_EL} cast into a linearized system of Helmholtz equations:
\begin{equation}\label{helmholtz}
    \left\{\begin{matrix}
\Delta f-\kappa^{2}_{f}f=0 \\
\Delta g-\kappa^{2}_{g}g=0,
\end{matrix}\right.
\end{equation}

where $\kappa_{f,g}^{2}=\mu^{2}_{f,g}-\omega^{2}_{f,g}$. Here, e.g. for the field fluctuation $f(r)$, one has the effective mass $\mu^{2}_{f}=\underset{r\rightarrow \infty }{\textup{lim} } \frac{1}{2}\frac{\partial^2 \mathcal{U}}{\partial f^2}$ and $\omega^{2}_{f}(\omega_{1},\omega_{2})=\underset{r\rightarrow \infty }{\textup{lim} } \frac{1}{2}\frac{\partial^2 \mathcal{L}_{0}}{\partial f^2}$, which is a binary quadratic form of $\omega_{1}$ and $\omega_{2}$  (see Sec.~\ref{sec:lagrangian_form}).\par
The solutions of \eqref{helmholtz} are exponentially decaying in space when $\kappa_{f,g}^{2}>0$, so for localized solitons to exist the system of two inequalities must hold (see \cite{AANS} as an example), yielding an ``upper bound" for frequencies
\begin{equation}\label{upper_bound}
    \mho_{+}\!\left ( \mu \right ) =\left\{ \left ( \omega_{1},\omega_{2} \right )\mid\omega^{2}_{f}\left ( \omega_{1},\omega _{2}\right )<\mu^{2}_{f}\;\text{ and }\;\omega^{2}_{g}\left ( \omega_{1},\omega _{2}\right )<\mu^{2}_{g}\right\}.
\end{equation}
\par
There is also another class of solutions, free particles or waves that oscillate near the vacuum state of the field, which are the case of either $\kappa_{f}^2<0$ or $\kappa^{2}_{g}<0$ or both. In the chosen ansatz, (see Sec.~\ref{sec:ansatz}), such waves may be expressed as
\begin{equation}
    \left\{\begin{matrix}
f\left ( r\rightarrow\infty  \right )\propto \frac{1}{\sqrt{r}}e^{\mathit{i}\left |  \kappa_{f}\right|r} \\
g\left ( r\rightarrow\infty  \right )\propto \frac{1}{\sqrt{r}}e^{\mathit{i}\left |  \kappa_{g}\right|r}
\end{matrix}\right.
\end{equation}
The lowest energy solution of such a kind is obtained by taking infrared limit of both waves, i.e. solving the system of algebraic equations

\begin{equation}
\left\{\begin{matrix}
\omega^{2}_{f}\left ( \omega_{1},\omega _{2}\right )=\mu^{2}_{f} \\
\omega^{2}_{g}\left ( \omega_{1},\omega _{2}\right )=\mu^{2}_{g}.
\end{matrix}\right.
\end{equation}

With the roots denoted as $\omega_{1}=m_{1}$ and $\omega_{2}=m_{2}$, it can be now seen from \eqref{E_constr}, that the absolute stability condition takes form
\begin{equation}\label{free_wave_stab}
E< E_{free}=\vec{m}\cdot\vec{Q}=m_{1}Q_{1}+m_{2}Q_{2}.
\end{equation}
Furthermore, since both upper \eqref{upper_bound} and lower bounds \eqref{lower_bound} are now defined, one may express
\begin{equation}\label{intersection_domain}
\left ( \omega_{1},\omega_{2} \right )\in\mho\left ( \mu \right ) =\mho_{-}\!\left ( \mu \right )\cap \mho_{+}\!\left ( \mu \right ),
\end{equation}
which is the domain of existence of localized stable solitonic solutions in frequencies space, see  the visualized case for potential \eqref{U_2_vac} in Ref.~\cite{AANS}.

\section{Numerical study \label{sec:numerical_study}}

\subsection{Methods \label{subsec:methods}}

Despite being a more natural description, Hamiltonian approach contains several serious obstacles when studied numerically. The biggest of them is the impossibility to formulate it in terms of ODEs, but only as IDEs, since the charges are integral quantities over physical space $\mathbb{R}^2$.
\par
Therefore, as it is described in Sec.~\ref{sec:legendre}, an optimizational approach for energy functional was adopted \cite{AAGNS, Hale}. All the results presented below were obtained with a 1D conjugate gradient method and also reproduced with a 1D simulated annealing method;
\par
Calculations were performed on a grid of 1000 nodes with an interval of $r\in \left [ 0, 30 \right ]$ so that the lattice spacing $ \Delta r=0.03$.

\subsection{Topological sectors \label{subsec:topological_sectors}}

\subsubsection{Potentials' symmetries
\label{subsubsec:potentials_symm}}

Despite apparent complexity of numerical study due to the model's abundance of degrees of freedom in the form of parameters \eqref{K_Omega} and boundary conditions of \eqref{full_EL}, it actually may be shown that by considering symmetric potentials, e.g. \eqref{U_2_vac} and \eqref{U_3_vac}, studying most of these solitonic configurations may be redundant due to their equivalence.
\par
Let us consider thee-vacuum potential \eqref{U_3_vac}. As it possesses $D_{3}$ symmetry, one can represent the action of elements $S$ and $R$ on homogeneous coordinate $Z$ using dual description \eqref{n_def}. Since the matrices of $S_{3}$ natural permutational representation are orthogonal, it actually may be shown that $D_{3} \cong  S_{3}\subset \textup{U}\! \left ( 3 \right )$, so that the corresponding actions on $Z$ are:
\begin{subequations}
\begin{align}
\begin{split}\label{Z_reflection}
 \left ( Z_{1},Z_{2},Z_{3} \right )^{T}\xrightarrow{\;S\;}\left ( Z_{2},Z_{1},Z_{3}  \right )^{T}
 \end{split}\\
 \begin{split}\label{Z_rotation}
\left ( Z_{1},Z_{2},Z_{3} \right )^{T}\xrightarrow{\;R\;}\left ( Z_{2},Z_{3},Z_{1}  \right )^{T}.
 \end{split}
\end{align}
\end{subequations}

Thus, acknowledging the above symmetries of \eqref{LagrangianZ1}, each of the solutions obtained in the ansatz \eqref{factor}-\eqref{X_profile_funs} 
\begin{equation}\label{explicit_ansatz}
    Z=\left ( \textup{cos}\! \left ( F \right ),\,
\textup{sin} \! \left ( F \right ) \textup{cos} \! \left ( G \right )e^{\textit{i}(k_{1}\theta+\omega_{1}t)},\,
\textup{sin} \! \left ( F \right ) \textup{sin} \! \left ( G \right )e^{\textit{i}(k_{2}\theta+\omega_{2}t)} \right )^{T}
\end{equation}
with "standard" boundary conditions
\begin{equation}\label{std_BC}
    F|_{r=0}=0;\; G|_{r=0}=\pi/2 \;\text{ and }\;F|_{r=\infty}=\pi /2;\; G|_{r=\infty}=0
\end{equation}
has the multiplicity of $|D_{3}|=|S_{3}|=6$ and, therefore, exactly this number of solutions with the same densities of EMT \eqref{EMT} and charges \eqref{charge_both_density} exists.
\par
That can be demonstrated on an example. Let us consider the following two sets of parameters: 

\begin{equation*}
    \textup{Set}_{1}\equiv \begin{Bmatrix}
\omega_{1}=1.827,\;\omega_{2}=2.003 \\
k_{1}=2,\;k_{2}=1 \\
F(0)=0,\;F(\infty)=\frac{\pi}{2}, \\
G(0)=\frac{\pi}{2},\;G(\infty)=0
\end{Bmatrix} \text{ and }
    \textup{Set}_{2}\equiv \begin{Bmatrix}
\omega'_{1}=2.003,\;\omega'_{2}=1.826 \\
k'_{1}=1,\;k'_{2}=2 \\
F(0)=0,\;F(\infty)=\frac{\pi}{2}, \\
G(0)=0,\;G(\infty)=\frac{\pi}{2}
\end{Bmatrix}
\end{equation*}

\begin{figure}[t]
	\centering
	\adjincludegraphics[trim={{.0\width} {.0\width} {.0\width} {.0\width}},clip, width=0.34\textwidth]{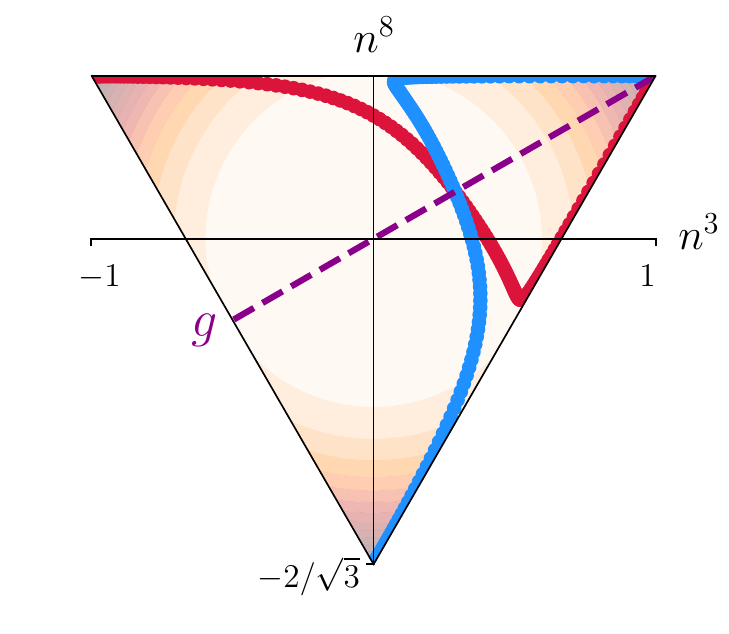}%
	\adjincludegraphics[width=0.45\textwidth]{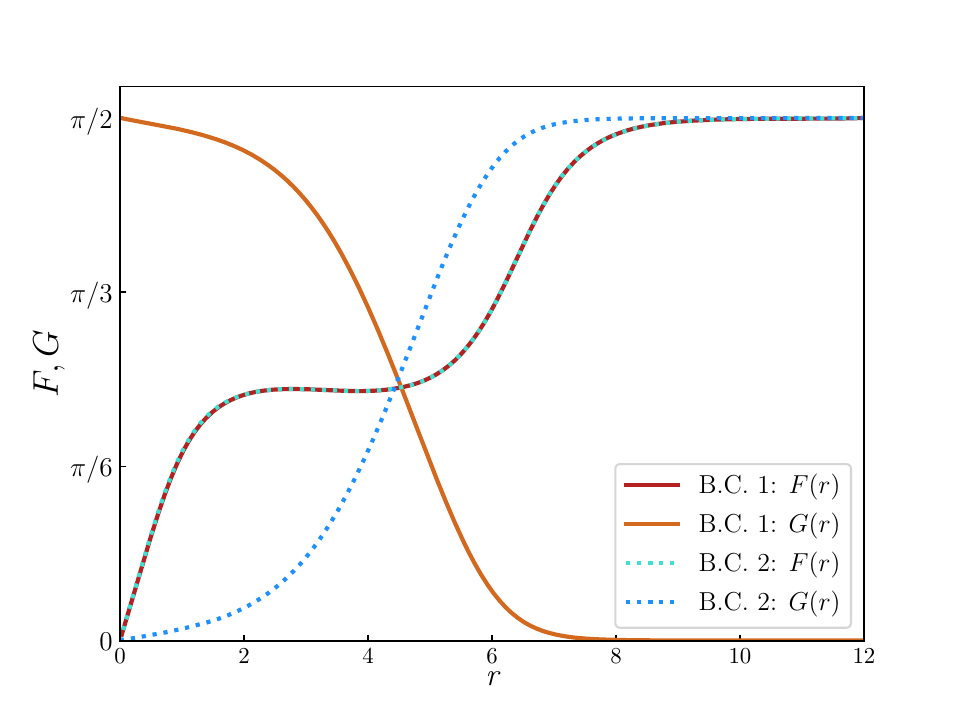}%
	\caption{Symmetry of the solution of the three-vacuum potential under global $D_{3}$ group action $g=S\boldsymbol{\cdot}R$; displayed on the toric diagram of $\mathbb{C}P^2$ (left plot) and on the physical space (right plot).}
	\label{symmetry_pic}
\end{figure}

These two sets are equivalent up to the transform $\textup{Set}_{1}\xrightarrow{S\boldsymbol{\cdot}R}\textup{Set}_{2}$ and actually produce solutions with exactly the same mechanical and topological characteristics, namely $E/4\pi=42.06$, $Q_{1}/4\pi=Q_{2}/4\pi=10.00$, and $Q_{\textup{top}}=1$. The profile functions are also the same up to the $G_{2}=\textup{arcsin}(\textup{cos}(G_{1}))$ transformation; see Fig.~\ref{symmetry_pic}.

To sum up, for any action of $g\in S_{3}$ on the ansatz Z \eqref{factor} it may once again be reverted to the form \eqref{explicit_ansatz}. Indeed:

\begin{equation*}
     gZ'=\begin{pmatrix}
X'_{a}e^{\textit{i}\left ( k'_{a}\theta+\omega'_{a}t \right )} \\
X'_{b}e^{\textit{i}\left ( k'_{b}\theta+\omega'_{b}t \right )} \\
X'_{c}e^{\textit{i}\left ( k'_{c}\theta+\omega'_{c}t \right )}
\end{pmatrix}\text{ mod U(1)}=
\begin{pmatrix}
X'_{a} \\
X'_{b}e^{\textit{i}\left ( (k'_{b}-k'_{a})\theta+(\omega'_{b}-\omega'_{a})t \right )} \\
X'_{c}e^{\textit{i}\left ( (k'_{c}-k'_{a})\theta+(\omega'_{c}-\omega'_{a})t \right )}
\end{pmatrix}=Z,
\end{equation*}

where $k'_{b}-k'_{a}=k_{1}$, $k'_{c}-k'_{a}=k_{2}$, $\omega'_{b}-\omega'_{a}=\omega_{1}$ and $\omega'_{c}-\omega'_{a}=\omega_{2}$. Similarly, $X'_{a}, X'_{b} \text{ and } X'_{c}$ are equated to $X_{1}, X_{2} \text{ and } X_{3}$, respectively. 
\par
All the above reasoning may, in a similar fashion, be applied to the symmetry properties of the potential \eqref{U_2_vac}. In this case, the reflection symmetry $D_{1}\cong \mathbb{Z}_{2}$ will yield a multiplicity of two for all the solutions. 
\par
It is also worth mentioning that all the equivalent solutions belong to the same topological sectors, since it is clear that \eqref{topChargeIntegral} is invariant under global $U(3)$ action.

\subsubsection{Analyticity in zero
\label{subsubsec:analiticity}}

Beyond the stability conditions (see Sec.~\ref{sec:existence_and_stability}), there is one more factor that may predetermine the existence of solitons; it concerns the analyticity of solutions in zero.
As the study revealed, the behaviour of the profile functions in zero depends only on quantum numbers and does not depend on potential.

Let us consider the equations of motion \eqref{full_EL} with boundary conditions \eqref{std_BC}. In the neighbourhood of zero for a perturbation of the form $F(r)=0+\epsilon\; f(r)$, $|\epsilon|\ll1$ one may obtain a Cauchy-Euler equation in the first order of $\epsilon$ smallness
\begin{equation*}
    r^{2}\, \frac{\partial ^{2} f}{\partial r^{2}}+r\, \frac{\partial  f}{\partial r}-k_{2}^{2}\ f=0,
\end{equation*}
which has a general solution 
\begin{equation*}
    f\left ( r \right )=\left\{\begin{matrix}
C_{1}r^{\left|k_{2} \right|}+C_{2}r^{-\left|k_{2} \right|} & k_{2}\neq 0  \\
C_{1}+C_{2}\textup{ln}(r) & k_{2}=0. \\
\end{matrix}\right.
\end{equation*}
Assuming $k_{2}\neq0$ and function being regular in zero, only one fitting particular solution remains:
\begin{equation}\label{sol_f}
f\left ( r \right )=C\, r^{|k_{2}|}.
\end{equation}
\par
Similarly, for the $G$ function a perturbation $G(r)=\frac{\pi}{2}-\epsilon\; g(r)$ is introduced, which yields the following equation 
\begin{equation}\label{lin_g_raw}
     r^{2} f \left(4\frac{\partial f}{\partial r}\,\frac{\partial g }{\partial r}+f\; \frac{\partial^2 g}{\partial r^2}\right)+ r f^{2}\, \frac{\partial g}{\partial r}-\left ( k_{1}^{2}+k_{1}k_{2}-2k_{2}^{2} \right ) f^{2} g=0.
\end{equation}
Substituting \eqref{sol_f} into \eqref{lin_g_raw} yields another Cauchy-Euler equation:
\begin{equation}\label{lin_g}
     r^{2}\, \frac{\partial ^{2} g}{\partial r^{2}}+\left ( 1+4\left| k_{2}\right| \right )\, r\, \frac{\partial g}{\partial r}-\left ( k_{1}^{2}+k_{1}k_{2}-2k_{2}^{2} \right ) g=0
\end{equation}

with a solution

\begin{equation*}
    g\left ( r \right )=\left\{\begin{matrix}
C_{1}r^{\alpha_{+} }+C_{2}r^{\alpha _{-}} & k_{1}\neq k_{2} \\
C_{1}+C_{2}r^{-4|k_{2}|} & k_{1}= k_{2}, \\ 
\end{matrix}\right.
\end{equation*}

where $\alpha _{\pm }=-2\left|k_{2} \right|\pm \sqrt{k_{1}^{2}+k_{1}k_{2}+2k_{2}^{2}}$. Since only positive $k_{1,2}$ are considered, one is restricted to a narrower subset of particular solutions:

\begin{equation}\label{sol_g}
    g\left ( r \right )=\left\{\begin{matrix}
C\, r^{\alpha_{+} } & k_{1}> k_{2} \\
C & k_{1}= k_{2}, \\ 
\end{matrix}\right.
\end{equation}

The case of $k_{1}=k_{2}$ for the function $G(r)$ may and will be regarded as Neumann boundary condition in zero. Indeed, such a degree of freedom does not violate the vacuum in $r=0$ since it is entirely determined by the function $F(r)$.

\begin{figure}[t]
	\centering
	\includegraphics[width=0.5\textwidth]{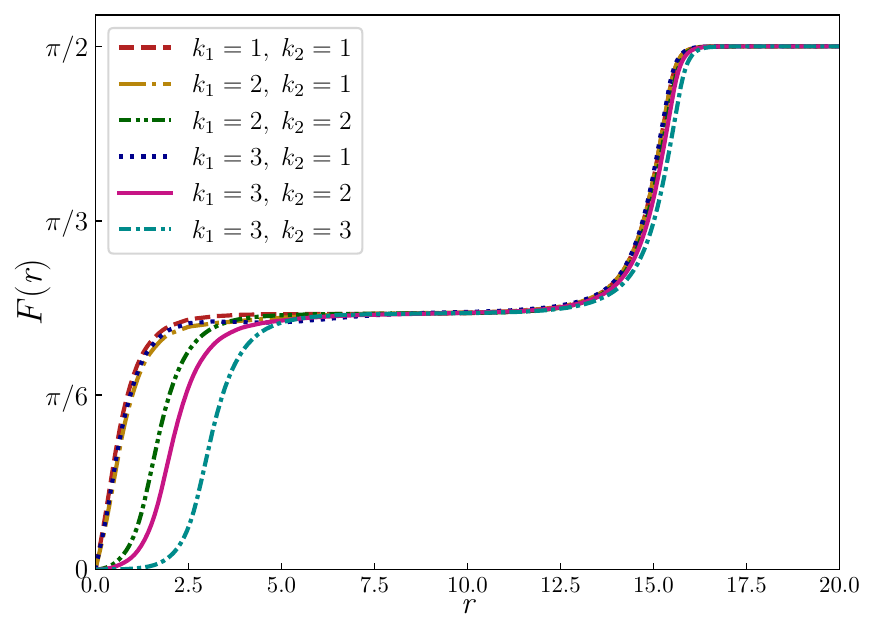}%
	\includegraphics[width=0.5\textwidth]{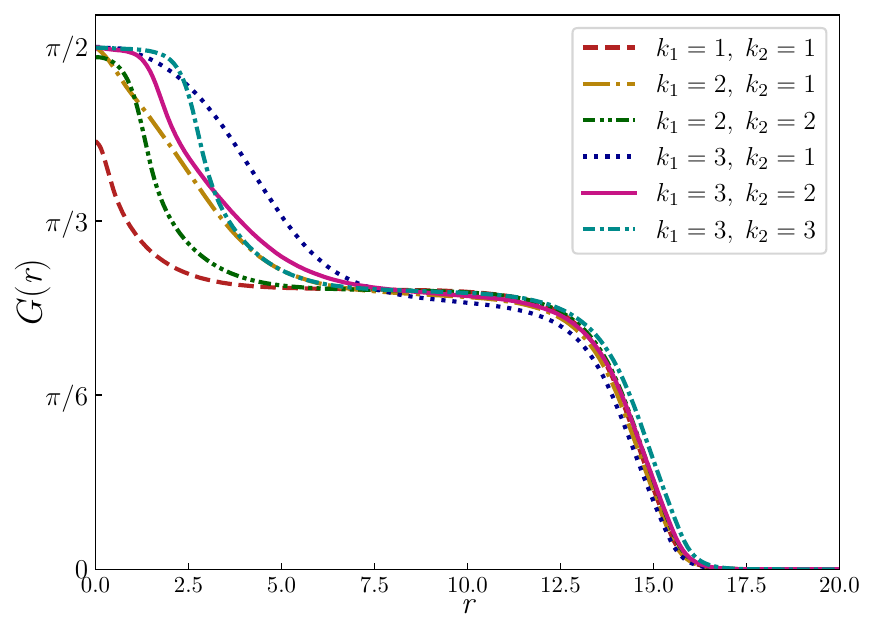}
	\caption{Profile functions $F(r)$ (left) and $G(r)$ (right) of a solitonic solutions with $\mu^{2}=1.0$ and  $Q_{1}/4\pi=Q_{2}/4\pi=47.5$ for various quantum numbers configurations.}\label{profiles_fgq1q2}
\end{figure}

\begin{figure}[t]
	\centering
	\includegraphics[width=0.5\textwidth]{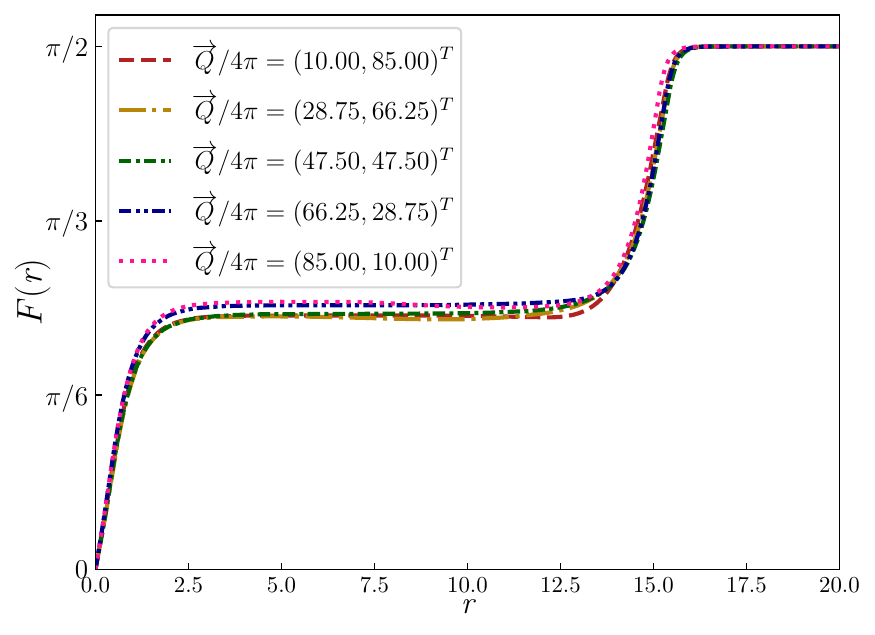}%
	\includegraphics[width=0.5\textwidth]{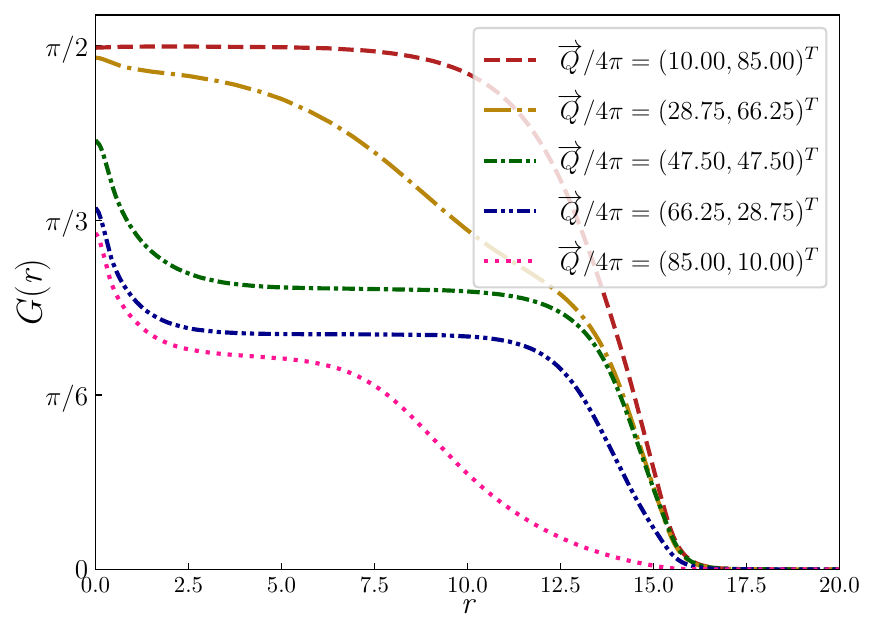}
	\caption{Profile functions $F(r)$ (left plot) and $G(r)$ (right plot) of various solitonic solutions with parameters $\mu^{2}=1.0$ and $k_{1}=k_{2}=1$ for different charges $(Q_{1}+Q_{2})/4\pi=95.0$.}\label{profiles_k1k2_1}
\end{figure}

\begin{figure}[t]
	\centering
	\includegraphics[width=0.5\textwidth]{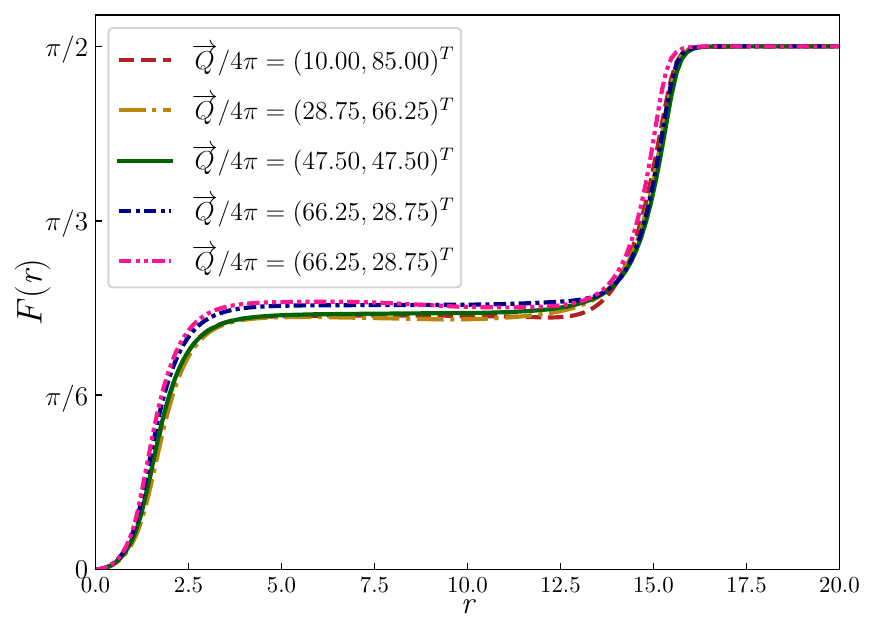}%
	\includegraphics[width=0.5\textwidth]{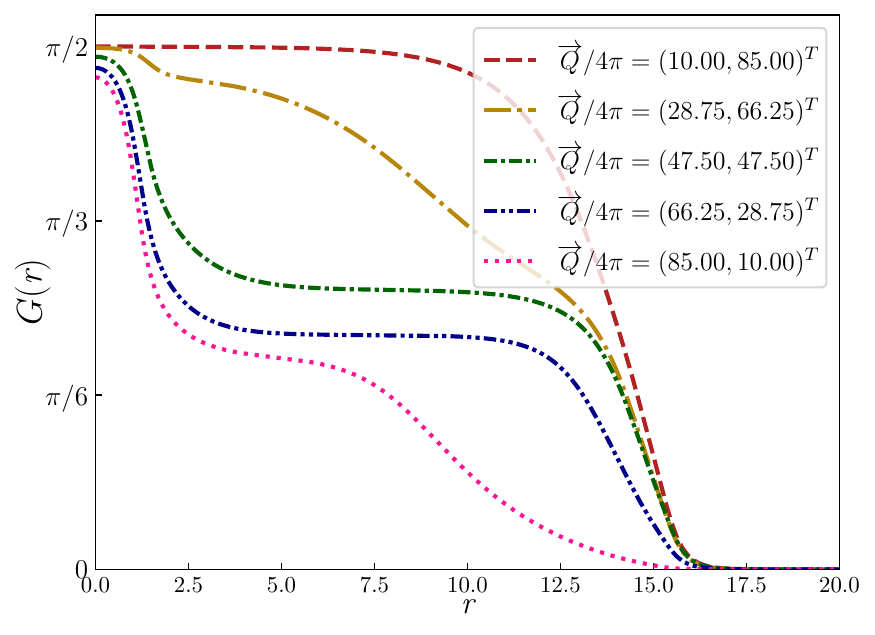}
	\caption{Profile functions $F(r)$ (left plot) and $G(r)$ (right plot) of various solitonic solutions with parameters $\mu^{2}=1.0$ and $k_{1}=k_{2}=2$ for different charges $(Q_{1}+Q_{2})/4\pi=95.0$.}\label{profiles_k1k2_2}
\end{figure}

Overall, for positive $k_{1,2}\in \left\{ 1,2,3\right\}$ the following set of solutions can be found: see Table~\ref{tab:solutions_zero} and Figures \ref{profiles_fgq1q2}, \ref{profiles_k1k2_1} and \ref{profiles_k1k2_2}

\begin{table}[!htb]
\centering
\caption{Analytical asymptotics for the profile functions in $r=0$ for various $k_{1,2}$.}
\label{tab:solutions_zero}
\begin{tblr}{
  cells = {c},
  hlines,
  vline{2} = {-}{0.075em},
  vline{3-4} = {-}{},
  hline{1,5} = {1}{-}{0.095em},
  hline{1,5} = {2}{-}{0.095em},
  hline{2} = {-}{0.075em},
}
\diagbox{$k_{1}$}{$k_{2}$} & 1  & 2  & 3  \\
1                        & $ \left\{\begin{matrix}
f\propto r \\
g\propto C
\end{matrix}\right.$ & -  & -   \\ 
2                        &  $ \left\{\begin{matrix}
f\propto r \\
g\propto r^{0.828}
\end{matrix}\right.$ &  $ \left\{\begin{matrix}
f\propto r^{2} \\
g\propto C
\end{matrix}\right.$ & -   \\
3                        &  $ \left\{\begin{matrix}
f\propto r \\
g\propto r^{1.742}
\end{matrix}\right.$ &  $\left\{\begin{matrix}
f\propto r^{2} \\
g\propto r^{0.796}
\end{matrix}\right.$  &  $ \left\{\begin{matrix}
f\propto r^{3} \\
g\propto C
\end{matrix}\right.$  \\
\end{tblr}
\end{table}

\subsection{Dynamical characteristics \label{subsec:dynamical_characts}}
\begin{figure}[t]
	\centering
	\includegraphics[width=0.5\textwidth]{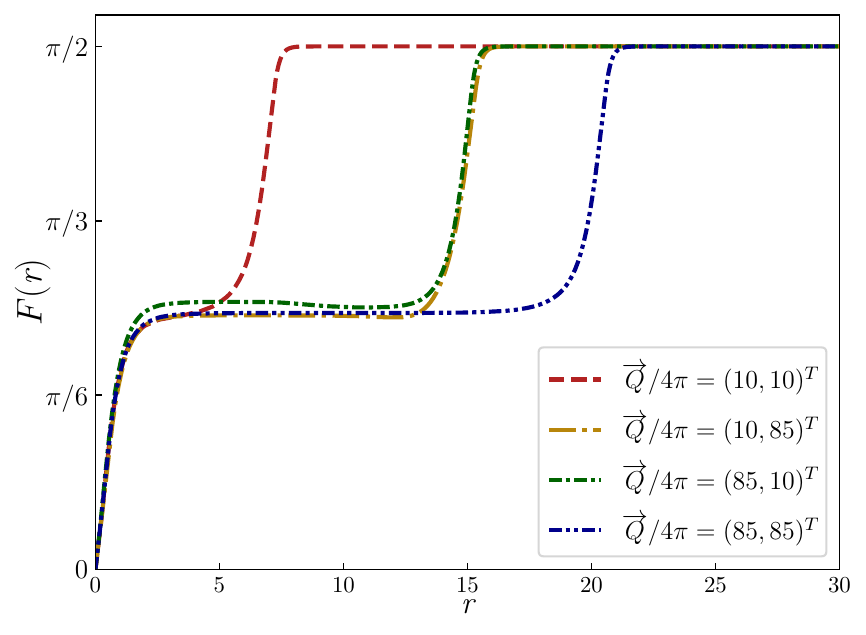}%
	\includegraphics[width=0.5\textwidth]{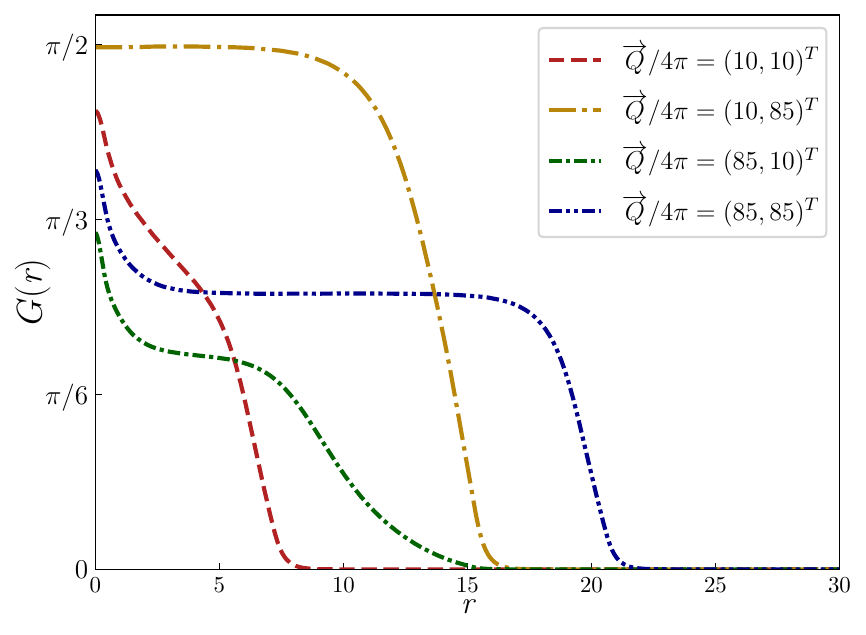}\\
	\includegraphics[width=0.5\textwidth]{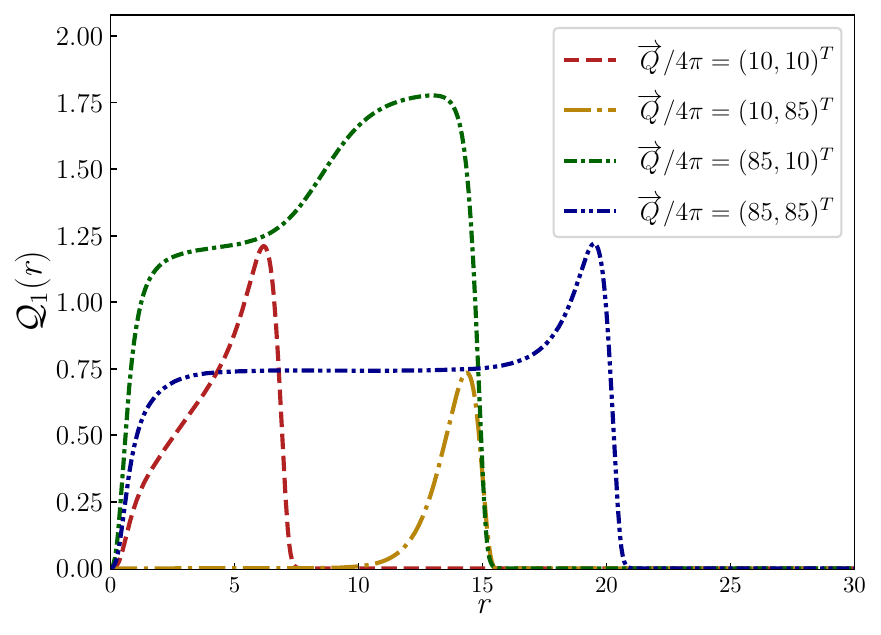}%
	\includegraphics[width=0.5\textwidth]{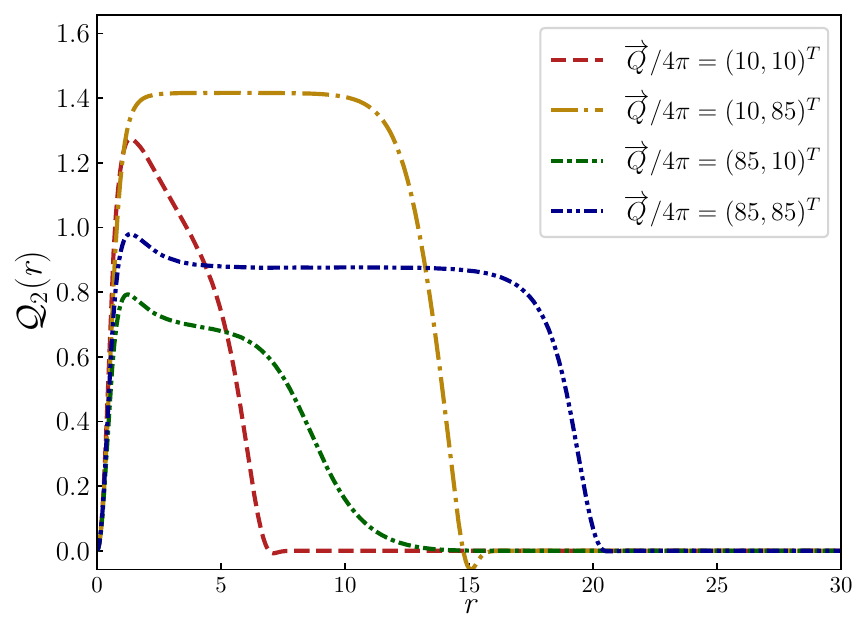}
	\caption{Profile functions $F(r)$ (upper left) and $G(r)$ (upper right) together with charge densities $\mathcal{Q}_{1}(r)$ (lower left) and $\mathcal{Q}_{2}(r)$ (lower right) for solitonic solutions with $\mu^{2}=1.0$ and $k_{1}=k_{2}=1$; various total charge configurations. Two-vacuum potential \eqref{U_2_vac}.}\label{profiles_2v_fgq1q2}
\end{figure}

\begin{figure}[t]
	\centering
	\includegraphics[width=0.5\textwidth]{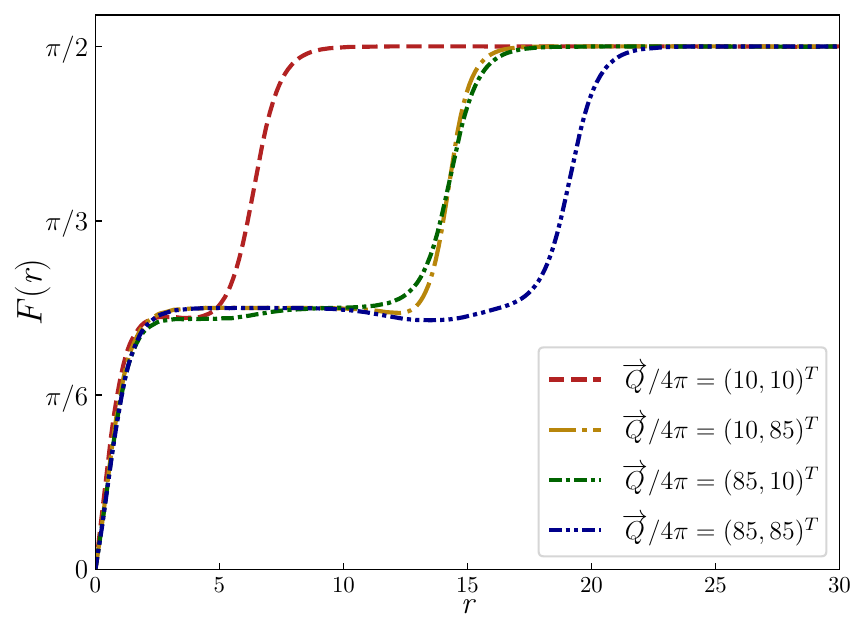}%
	\includegraphics[width=0.5\textwidth]{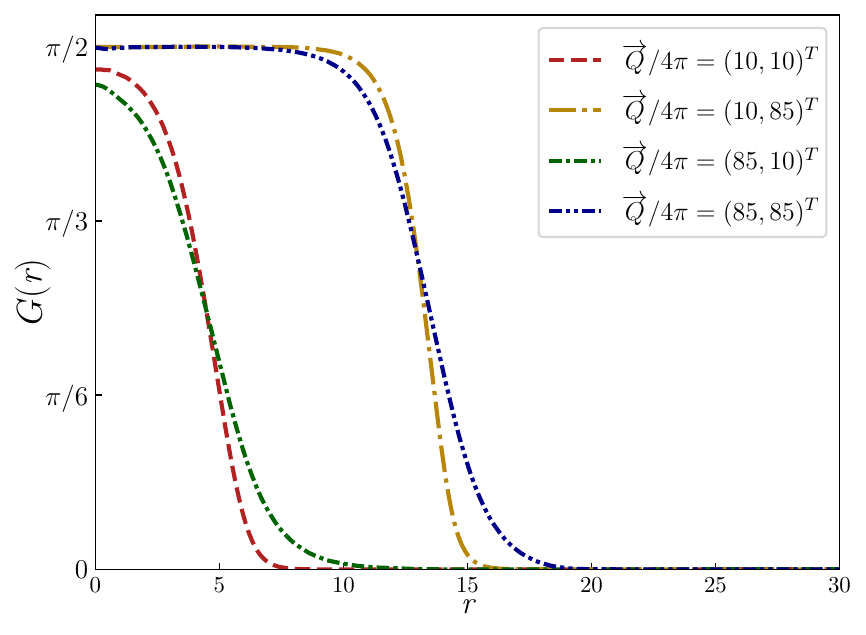}\\
	\includegraphics[width=0.5\textwidth]{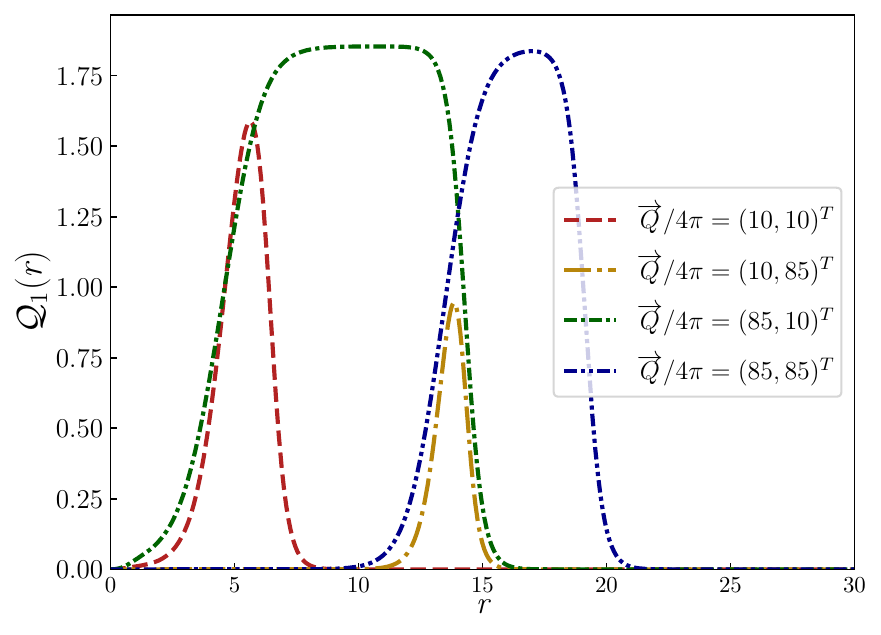}%
	\includegraphics[width=0.5\textwidth]{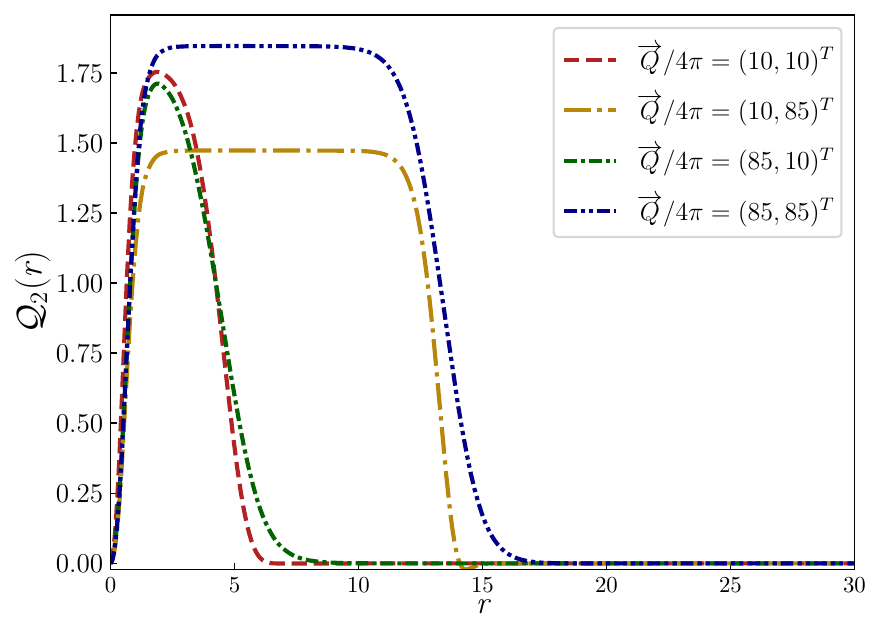}
	\caption{Profile functions $F(r)$ (upper left) and $G(r)$ (upper right) together with charge densities $\mathcal{Q}_{1}(r)$ (lower left) and $\mathcal{Q}_{2}(r)$ (lower right) for solitonic solutions with $\mu^{2}=1.0$ and $k_{1}=k_{2}=1$; various total charge configurations. Three-vacuum potential \eqref{U_3_vac}.}\label{profiles_3v_fgq1q2}
\end{figure}

First of all, let us examine the structure of single solitons with respect to their Noether charges, see Fig.~\ref{profiles_2v_fgq1q2} and Fig.~\ref{profiles_3v_fgq1q2}.
\par
One may notice that distinctive plateau corresponding to effective potential's minimum can be observed in the case of $F(r)$ profile function, when in the case of $G(r)$ it is more variable in its extent, height or completely absent. Also, comparing the general patterns of profile functions for both the potentials, it may be observed that they are shared between $F(r)$ and differ visibly for $G(r)$.

In the same way differences may be seen between Noether charges' densities: $\mathcal{Q}_{1}$ has much sharper edge, while $\mathcal{Q}_{2}$ descends more gently.

\begin{figure}[t]
	\centering
	\includegraphics[width=0.5\textwidth]{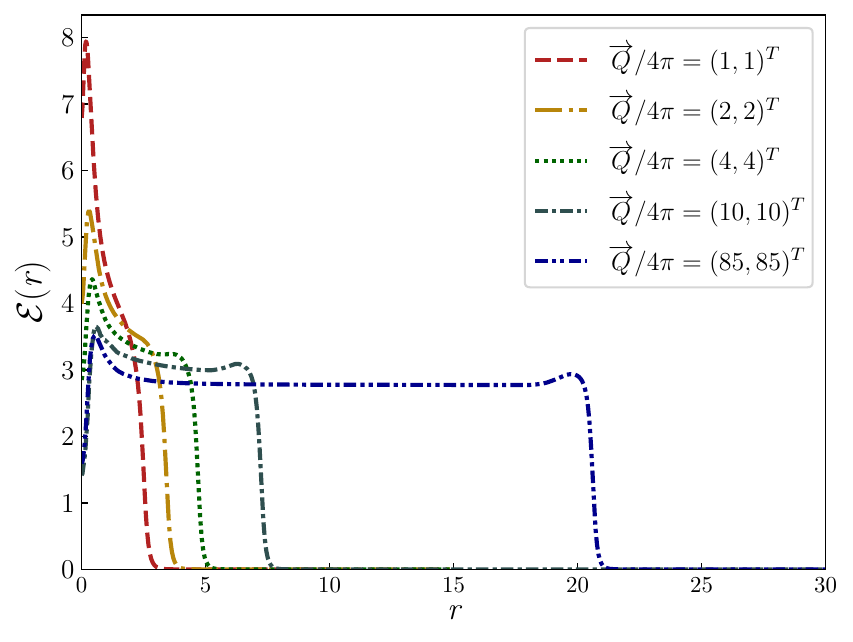}%
	\includegraphics[width=0.5\textwidth]{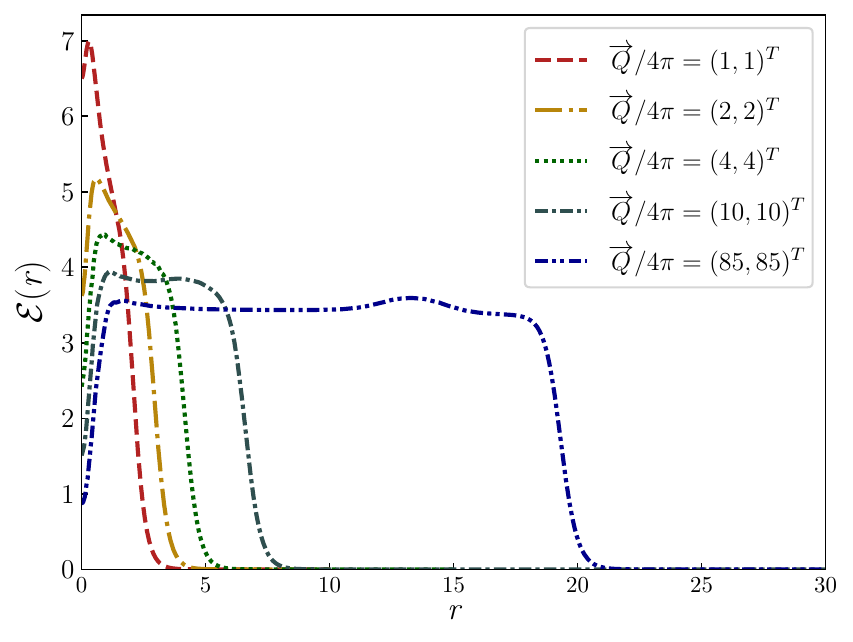}%
	\caption{Energy densities $\mathcal{E}(r)$ of solitonic solutions with parameters $\mu^{2}=1.0$ and $k_{1}=k_{2}=1$ for different total charge configurations. Two-vacuum (left) and three-vacuum (right) potentials.}\label{profiles_en}
\end{figure}

As for the profiles of energy density with respect to total charges: a vast plateau associated with volume energy dominance tends to emerge when approaching thin-wall limit, see Fig.~\ref{profiles_en}. Both the potentials seem to share such a behaviour.
\par
Now lets examine how do solitons' integral characteristics behave as functions of Noether charges. 
\par
What can be noted about $\omega_{1}, \omega_{2}$ is that they have a pattern, similar to that of $U(1)$ Q-balls \cite{Ward, MareikeThesis}: a spike in the region of small charges, corresponding to the upper limit of frequencies and an asymptotic descent to some constant lower limit values. All the frequencies obtained for both the potentials appeared to lie within the corresponding domains \eqref{intersection_domain}.

\begin{figure}[t]
	\centering
	\adjincludegraphics[trim={{.165\width} {.105\width} {.195\width} {.17\width}},clip, width=0.5\textwidth]{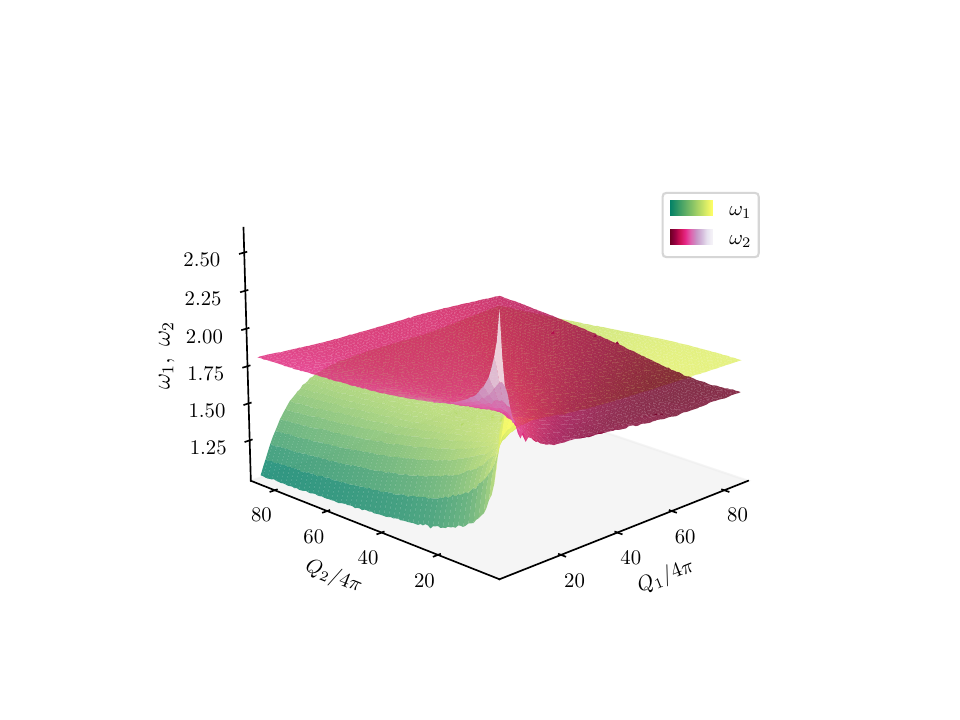}%
	\adjincludegraphics[trim={{.155\width} {.105\width} {.205\width} {.17\width}},clip, width=0.5\textwidth]{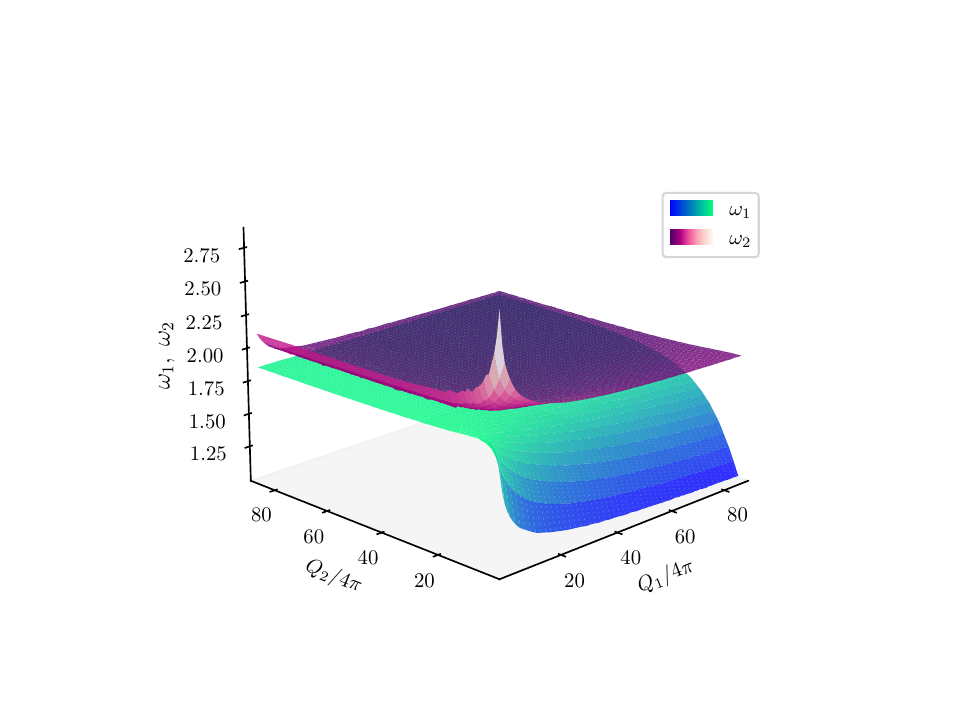}%
	\caption{Isorotational frequencies as functions of total Noether charges for $\mu^{2}=1.0$ and $k_{1}=k_{2}=1$. Lattice setup: 100$\times$100 points, charges interval $Q_{1},\, Q_{2}\in \left [ 10, 1100 \right ]$. Two-vacuum potential (left) and three-vacuum potential (right).}
	\label{freqs_surfs}
\end{figure}

\begin{figure}[t]
	\centering
	\adjincludegraphics[trim={{.175\width} {.105\width} {.195\width} {.17\width}},clip, width=0.5\textwidth]{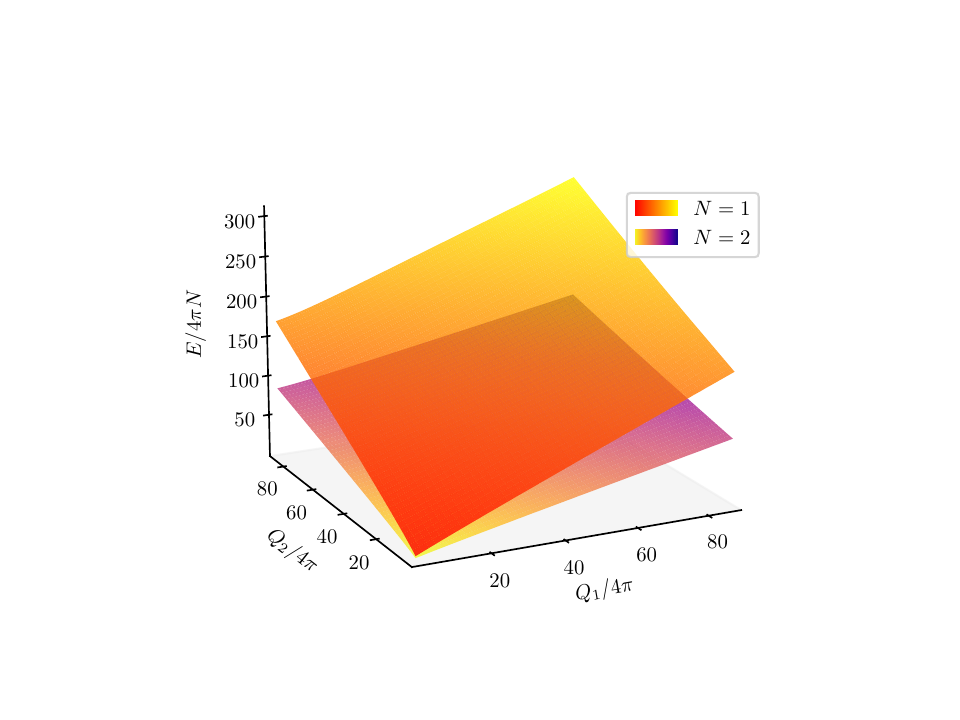}%
    \adjincludegraphics[trim={{.175\width} {.105\width} {.195\width} {.17\width}},clip, width=0.5\textwidth]{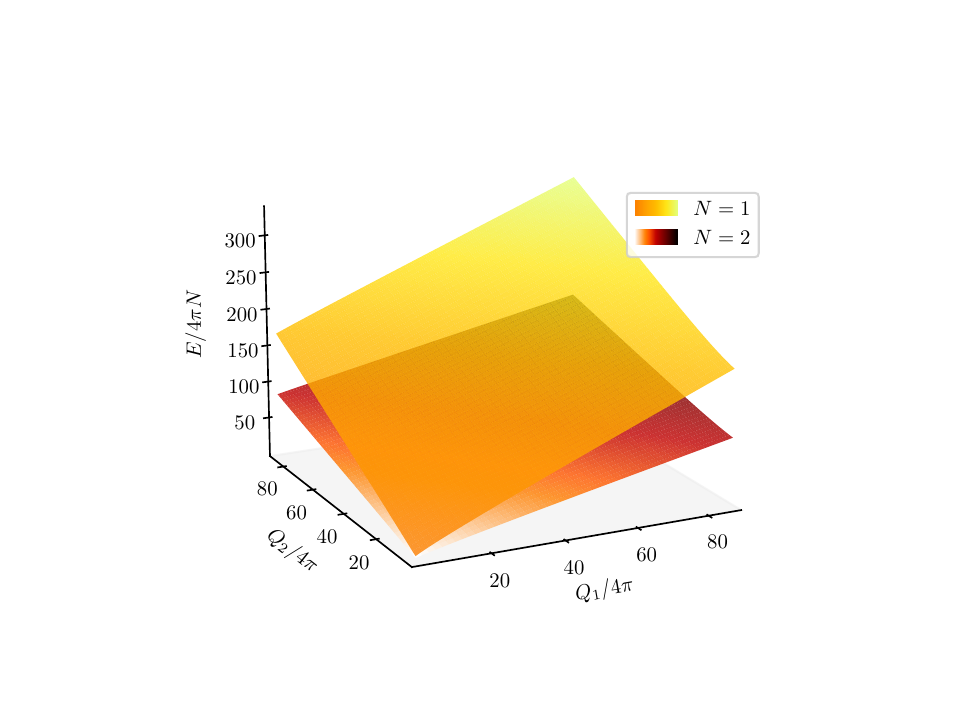}
	\caption{Total energy per soliton as a function of total Noether charges. Lattice setup: 100$\times$100 points, charges interval $Q_{1},\, Q_{2}\in \left [ 10, 1100 \right ]$. Quantum numbers are set as $k_{1}\!=\!k_{2}\!=\!1$ (topological charge $N=1$) and $k_{1}\!=\!2, \,k_{2}\!=\!1$ (topological charge $N=2$). Two-vacuum potential (left) and three-vacuum potential (right).
	}
	\label{en_surfs}
\end{figure}

Energy also shows behaviour similar to U(1) Q-balls: to a decent approximation there is a linear dependency on charges, same as for the one-parametric case \cite{Ward}. 
Regarding stability, the virial relation is satisfied up to ~$10^{-3}$ accuracy, see Table~\ref{tab:freqs_and_erg}.
Solutions are also stable towards the fission into plane waves. Indeed, for the potential \eqref{U_2_vac} one has the lowest free particles' parameters as 
$m_{1}=\sqrt{30}\approx 5.477$ and $m_{2}=\sqrt{\frac{15}{2}}\approx 2.739$, so with the Fig.~\ref{en_surfs} given, the inequality \eqref{free_wave_stab} seems to be satisfied. 
Similarly, for \eqref{U_3_vac} one has $m_{1}=\frac{4}{\sqrt{3}}\approx 2.309$ and two options for the second wave: either $m_{2}=\frac{8}{\sqrt{3}}\approx 4.619$ or $m_{2}=0$. In the former case the solitons are stable on the whole studied domain of $Q_{1,2}$, in the latter case they are quantum mechanically stable within the region of $E<\frac{4}{\sqrt{3}}Q_{1}$, see Fig.~\ref{en_surfs}.

\begin{table}[!htb]
\captionsetup{justification=raggedright,singlelinecheck=false}
\centering
\resizebox{\linewidth}{!}{%
\begin{talltblr}[caption = {Energies per particle as functions of quantum numbers and Noether charges.}, label={tab:freqs_and_erg}]{
  width = \linewidth,
  colspec = {Q[20]Q[83]Q[83]Q[83]Q[83]Q[83]Q[83]Q[83]Q[83]},
  rows = {abovesep=0.4pt,belowsep=0.4pt},
  cell{1}{1} = {c=2}{0.126\linewidth},
  cell{2}{1} = {r=12}{},
  cell{2}{2} = {r=4}{},
  cell{6}{2} = {r=4}{},
  cell{10}{2} = {r=4}{},
%  vline{3,5} = {1-10}{},
%  vline{2} = {2-10}{},
  hline{1,14} = {1}{-}{0.095em},
  hline{1,14} = {2}{-}{0.095em},
  hline{2} = {1-9}{},
hline{6} = {1-9}{},
hline{10} = {1-9}{},
}
                                              &                                     & $Q_{1}/4\pi$ & $Q_{2}/4\pi$ & $E/4\pi N$ & $L_{0}/4\pi N$ & $U/4\pi N$ & $\omega_{1}$ & $\omega_{2}$ \\
\begin{sideways}Quantum numbers\end{sideways} & $ \vec{\textbf{k}} \!  =\!  \left(\begin{matrix}
1 \\
1
\end{matrix}\right)$ &  10.00    &   10.00     &   39.98      & 17.96       &   17.97      &    1.707     &      1.885   \\
&      &   10.00     &   85.00      & 176.9       &   83.40      &   83.56     &      1.421& 1.795   \\
                                              &                                     &  85.00     &  10.00       &    173.8     &   83.78    &   83.90      &   1.788      &    1.559     \\
                                              &                                     &   85.00   &   85.00      & 297.4        &  143.6      &     143.9     &     1.654    &   1.725      \\
                                             & $ \vec{\textbf{k}} \!  =\!  \left(\begin{matrix}
2 \\
1
\end{matrix}\right)$ &  10.00    &   10.00     &   20.27      & 9.002       &   9.024      &    1.756     &      1.845   \\ 
&      &   10.00     &   85.00      & 88.47       &   41.67      &    41.77     &      1.423 & 1.794   \\ 
                                                &                                     &  85.00     &  10.00       &    87.67   &   41.96    &   42.01      &   1.792      &    1.551     \\   
                                                &                                     &   85.00   &   85.00      & 149.2       &  71.88      &     72.06     &     1.664    &   1.719     \\            
                                              & $ \vec{\textbf{k}} \!  =\!  \left(\begin{matrix}
2 \\
2
\end{matrix}\right)$ &  10.00    &   10.00     &   20.72      & 9.045       &   9.041      &    1.725     &      1.893   \\
&      &   10.00     &   85.00      & 89.43       &   41.77      &    41.77     &      1.422& 1.799   \\
                                              &                                     &  85.00     &  10.00       &    87.97     &   42.02    &   41.93      &   1.793      &    1.565     \\
                                              &                                     &   85.00   &   85.00      & 149.8        &  71.91      &     71.83     &     1.657    &   1.726       
\end{talltblr}
}
\end{table}

\subsubsection{Lower energy bound \label{subsubsec:lower_energy}}

When considering the total energy as a function of charges $E\left(Q_{1},Q_{2}\right)$ (see Fig.~\ref{en_surfs}), a clear plane-like pattern may be observed, along with a lack of visible dependence on the choice of a potential whatsoever. This pattern may actually be explained by a BPS-like bound \cite{Leese}.
\par

To derive it, the following inequality has to be simplified
\begin{equation*}
    \int \textup{d}^{2}x\bigl( (D_{j}Z)^{\dagger }\pm \textit{i}\, \varepsilon _{jk}(D_{k}Z)^{\dagger } \bigr)\bigl( D_{j}Z\mp \textit{i}\, \varepsilon _{jl}D_{l}Z \bigr)\geqslant0,
\end{equation*}
% where $\varepsilon _{jk}$ is absolutely antisymmetric $2\times2$ tensor with $\epsilon _{12}=1$.
Using \eqref{topChargeIntegral}, one may obtain
\begin{equation*}
   2 \!\int \textup{d}^{2}x \bigl((D_{j}Z)^{\dagger }D_{j}Z  \bigr)\geqslant 4\pi \left|Q_{\textup{top}} \right|.
\end{equation*}

Now, assuming that only stable solitonic configurations are considered, which is the case of this paper, the virial relation \eqref{virial} may be applied.
Adding twice the $L_{0}=U$ to the both sides of the inequality, one may complete total energy on the left and simplify \eqref{L0Q} on the right. The expression is now recast into the final form
\begin{equation}\label{D_BPS}
    E_{\textup{Stable}}\geqslant 4\pi \left|Q_{\textup{top}} \right|+\omega_{1} Q_{1}+\omega_{2} Q_{2},
\end{equation}
which is a combined Derrick-Bogomolny bound for stable Q-ball solutions. 
Although neither of the potentials \eqref{U_2_vac}, \eqref{U_3_vac} supports an exact analytical solutions for the saturation case, the above formula may be viewed as a good numerical estimate of total energy value. As the charges grow, the frequencies tend to their lower limit (see Fig.~\ref{freqs_surfs}), so that \eqref{D_BPS} can be regarded as a linear function of charges, i.e. a plane. Indeed, taking into account numerical results for potential \eqref{U_2_vac} in Fig.~\ref{en_surfs} and in Table~\ref{tab:freqs_and_erg}, one may observe decent quantitative and qualitative agreement, respectively, with such an approximation. 

\section{Conclusions \label{sec:conclusions}}

As a continuation of the previous study \cite{AANS}, this paper addressed a number of new problems.
\par 
Firstly, topological sectors were studied thoroughly (see Sec.~\ref{subsec:topological_sectors}). It was shown that different combinations of quantum numbers $k_{1,2}$ yield various behaviours for the profile functions in the centre of a soliton (see Table~\ref{tab:solutions_zero}). A unit charge sector, $Q_{\textup{top}}=1$, was given special attention, due to being omitted in the previous work. 
\par
As it was established, even without a suitable potential, the model may still possess a Bogomolny-like lower limit on energy \eqref{D_BPS}, but an assumption must be made that the solution is stable. The study shows such a bound to be saturated for both the potentials, \eqref{U_2_vac} and \eqref{U_3_vac}, see Fig.~\ref{en_surfs}. 
\par
The stability, as a key part of solitons' phenomenology, has been investigated. The $\mathbb{C}P^N$ possess exactly the same virial relation as a $U(1)$ model. A quantum stability condition is also similar, yet the number of decay channels is two now.
\par
One more notable thing is how charges are mapped onto another mechanical quantity, the angular momentum \eqref{angular_m}. It appeared to be classically quantized, just as for $U(1)$ Q-balls \cite{Radu}, with both the Noether charges' contributions being proportional to the corresponding quantum numbers. 
Thus, while the charges are observables individually, they are also projected onto a single measurable \cite{WOLFF} mechanical quantity via the non-injective mapping $f\!:\!\mathbb{R}^{2}\!\to\!\mathbb{R}$, establishing a connection between isorotations in the target space and rotation in the physical space.
\par
One can perceive these work as a preliminary for studying planar $\mathbb{C}P^N$ Q-solitons. Sections~\ref{sec:introduction}~-~\ref{sec:legendre} may be considered in general way, without adhering to the number and the form of particular maximal tori $U(1)^{N}$ generators, therefore yielding a hamiltonian formulation for any $\mathbb{C}P^N$, with $\mathbb{C}P^1$ case virtually being $SO(3)$ Q-balls \cite{MareikeThesis}.

\section{Acknowledgements \label{sec:acknowledgements}}

I would like to express my gratitude to Yuki Amari, Muneto Nitta and Yakov Shnir for problem formulation and valuable discussions.

\end{document}